\documentclass[runningheads]{llncs}
\pdfoutput=1

\AtBeginDocument{%
  \providecommand\BibTeX{{%
    \normalfont B\kern-0.5em{\scshape i\kern-0.25em b}\kern-0.8em\TeX}}}

\usepackage{array,longtable,makecell,multirow}
\usepackage{booktabs}
\usepackage{lscape}

\usepackage{algorithmic}
\usepackage{graphicx}
\usepackage{textcomp}
\usepackage{xcolor}
\usepackage{wrapfig}

\usepackage{ragged2e}

\usepackage{multirow}
\usepackage{hyperref} 
\usepackage[utf8]{inputenc}
\usepackage{textcomp}

\usepackage{comment}%for block comment

\usepackage{empheq}

\usepackage{breakcites} %To break long citations

\usepackage[normalem]{ulem}
\usepackage{soul} %for text highlighting 

\usepackage{longtable} %for long table

\usepackage[utf8]{inputenx}
\usepackage[T1]{fontenc}
\usepackage{color,soulutf8}

\makeatletter
\def\subsubsection{% 
  \@startsection
    {subsubsection}                 % type
    {3}                             % level
    {\parindent}                    % indent
    {2.5ex plus 1.5ex minus 1.5ex}  % beforeskip {0ex plus 0.1ex minus 0.1ex}
    {0.7ex plus .5ex minus 0ex}     % afterskip {0ex}
    {\normalfont\normalsize\itshape}% style
} 
\newcommand{\linebreakand}{%
  \end{@IEEEauthorhalign}
  \hfill\mbox{}\par
  \mbox{}\hfill\begin{@IEEEauthorhalign}
}

\makeatother

\usepackage[left=2.5cm, top=3.2cm, asymmetric]{geometry} %asymmetric is to similarly apply the margines to the odd and even pages 

\begin{document}

\title{Threat Modeling and Security Analysis of Containers: A Survey}

%%
%% The "author" command and its associated commands are used to define the authors and their affiliations.

%\renewcommand{\shortauthors}{Wong, Chekole, Ochoa and Zhou et al.}

\author{Ann Yi Wong\inst{1} \and Eyasu Getahun Chekole\inst{1} \and Mart\'in Ochoa\inst{2} \and Jianying Zhou\inst{1}}

\institute{Singapore University of Technology and Design, Singapore 487372, Singapore\\ 
\email{annyi\_wong@mymail.sutd.edu.sg}, \email{\{eyasu\_chekole, jianying\_zhou\}@sutd.edu.sg}
\and Department of Computer Science, ETH Zurich, 8092 Zurich, Switzerland\\
\email{martin.ochoa@inf.ethz.ch}
}
\maketitle

\begin{abstract}
 
% For a long time, 
 Traditionally, applications that are used in large and small enterprises were deployed on ``bare metal'' servers installed with operating systems. Recently, the use of multiple virtual machines (VMs) on the same physical server was adopted due to cost reduction and flexibility. Nowadays, containers have become popular for application deployment due to smaller footprints than the VMs, their ability to start and stop more quickly, and their capability to pack the application binaries and their dependencies/libraries in standalone units for seamless portability. A typical container ecosystem includes a code repository (e.g., GitHub) where the container images are built from the codes and libraries and then pushed to the image registry (e.g., Docker Hub) for subsequent deployment as application containers. However, the pervasive use of containers also leads to a wide-range of security breaches such as attackers stealing credentials, source codes and sensitive data from image registry and code repository, carrying out DoS attacks on application containers, and gaining root access to misuse the underlying host resources, among others. %In this paper, we %\revised{survey existing works on the container ecosystem and} 
 %apply a threat modeling framework, called STRIDE, to identify the vulnerabilities that exist in each system component, investigate potential threats and their consequences, and subsequently present existing mitigation strategies. Our paper is aimed to help researchers and practitioners to gain a deeper understanding of the threat landscape in containers and the state-of-the-art countermeasures. % some future research directions. 
 %revised
 In this paper, we first perform threat modeling on the containers ecosystem using the popular threat modeling framework, called STRIDE. Using STRIDE, we identify the vulnerabilities in each system component, and investigate potential security threats and their consequences. Then, we conduct a comprehensive survey on the existing countermeasures designed against the identified threats and vulnerabilities in containers. In particular, we assess the strengths and weaknesses of the existing mitigation strategies designed against such threats. We believe that this work will help researchers and practitioners to gain a deeper understanding of the threat landscape in containers and the state-of-the-art countermeasures. We also discuss open research problems, the research gaps and future research directions in containers security, which may ignite further research to be done in this area. 
\end{abstract}


%\maketitle

\section{Introduction}\label{introduction}

Many enterprises have started to deploy applications in containers. Some popular examples are Gmail, YouTube, Google Search \cite{google-a}, Netflix \cite{hall2018a}, and PayPal financial services \cite{armstrong2017a}, among others. %and many more. 
Running an application in a container allows its binaries, libraries, and other dependencies to be abstracted from the operating environment and hence be portable from a developer notebook to the on-prem data centre and the public cloud. Therefore, containerization allows an application to be deployed %\sout{faster,}
efficiently and scaled easily. Gartner, a leading research and advisory company in information technology and cybersecurity %and other fields, 
forecasts that 15\% of all applications will be running in containers by 2024, up from 5\% in 2020 \cite{williams2020a}.
Gartner also forecasts that 75\% of large enterprises globally will deploy production application in containers by 2022, up from less than 30\% in 2020 \cite{williams2020a}. The most widely employed container runtime is Docker at 79\% share of the market \cite{vizard2019a}.\\
\indent %Despite the many benefits they bring in to enterprises, 
Although containers are revolutionizing enterprises and other systems, %have also their security breaches and vulnerabilities they are also vulnerable to 
they also have several weaknesses and vulnerabilities that expose them to a wide-range of cyberattacks. A recent report \cite{barua2020a} revealed that about 51\% of around 4 million images in Docker Hub have exploitable vulnerabilities of which 0.16\% or 6,432 images had malicious software which were primarily cryptocurrency miner. The attackers could insert malicious images directly on misconfigured hosts \cite{barua2020a}, \cite{field2020a} or into Docker Hub due to the ease of pushing and pulling images to and from it without controls \cite{barua2020a}.  
%and images embedded with malware 
In another report \cite{cimpanu2019b}, %a cybersecurity group through their regular monitoring found that towards the end of 2019, a hacking group mass-scanned more than 59,000 IP networks to look for exposed Docker API endpoints. 
a cybersecurity team discovered through its regular monitoring that by the end of 2019, a hacker group scanned more than 59,000 IP networks on a large scale to find exposed Docker API endpoints. Most containers are also configured with default network settings, %which enables remote connections to be established easily.
making it easy to establish remote connections . This was discovered by TeamTNT (a cybercrime group) and used it as a backdoor to run crypto-mining malware on the underlying system to generate cryptocurrencies \cite{vizard2020a}. As of the date of this paper, there are 428 container related security vulnerabilities listed in MITRE CVE \cite{mitre2021a}. \\
\indent Several real-world cyberattacks have also been reported on containers. In 2018, attackers hacked into Tesla’s container orchestration console of Kubernetes and installed crypto-mining software to mine cryptocurrency using its cloud computing resources \cite{seals2018a}. %Meanwhile, 
Consequently, the U.S. government National Security Agency (NSA) also alerted %the industry 
industries over a foreign-based cybercrime group APT28's massive attacks on containers %which are run 
that run in Kubernetes clusters \cite{nichols2021a}. In 2019, other attackers hacked into Docker Hub and gained access to usernames and passwords of 190,000 user accounts \cite{matthews2019a}. %When an exposed Docker host was detected, the attackers would use it to spin a container which would 
An attacker can then use the compromised Docker instance as a backdoor to spin the container, which will install the XMRRig cryptocurrency miner for illegal mining \cite{cimpanu2019b}. There were also many other critical attacks that had been launched on containers and their subsystems \cite{townsend2021a}, \cite{jarvis2020a}, \cite{morag2020a}, \cite{remillano2020a}, \cite{sagi2019a}, \cite{chako2020a}, \cite{martin2018a}, \cite{kerner2019a}, \cite{liu2020a}, \cite{shen2020a}, \cite{luo2016a}, \cite{redlockcsi2018a}, \cite{martin2018a}, \cite{gao2019a}, \cite{pavisic2019a}. 
These and other real-world examples show how security is a critical concern in container systems, beyond the conventional IT systems.\\ 
\indent To alleviate the security concerns, several research works have been done on containers security, some focusing on vulnerability analysis \cite{combe2016a}, \cite{duarte2018a}, \cite{wist2020a}, \cite{flauzac2020a}, \cite{sultan2019a}, \cite{lin2018a}, \cite{burns2021a}, \cite{mitre2021b}, and others on mitigation strategies \cite{oh2017a}, \cite{redhat2021a}, \cite{bui2015a}, \cite{docker2021p}, \cite{docker2021i}, \cite{iradier2021a}, \cite{docker2021a}, \cite{sun2018a}, \cite{bhat2018a}, \cite{brady2020a}, \cite{souppaya2020a}, \cite{martin2018a},\cite{cis2021a}, \cite{docker2021m}, \cite{lei2017a}. However, most of the related works only focus on a specific vulnerability, threat, use-case or subsystem of containers. Hence, they do not provide a comprehensive security analysis on the entire container ecosystem (spanning image creation to distribution processes). In addition, most of the existing mitigation strategies already have certain flaws and limitations. For example, recent studies revealed that the existing Linux-based mitigation strategies used in containers, such as cgroups, namespaces and capabilities, are subjected to attacks resulting in resources exploitations and denials of services \cite{gao2019a}, \cite{martin2018a}. Furthermore, some are probably outdated and may not reflect the latest threat landscape as shown in the example of \cite{bui2015a}, which suggests that the Docker container is fairly secure with the default configuration but it is in fact exploitable in today's context \cite{stoler2021a}. Therefore, the existing works might not provide a comprehensive security analysis and state-of-the-art information on the security landscape of the containers ecosystem.\\  
%
%\revised{
\indent In this work, we make a systematic %review and analysis %and study on 
and comprehensive survey on the security of containers, covering vulnerabilities, threats, threat consequences and existing mitigation strategies, to provide a comprehensive and state-of-the-art information on the security landscape of containers. To be able to specify the scope of our survey and map existing literature, we first perform threat modeling on the containers ecosystem. In particular, we study the threat landscape of the containers supply chain process -- spanning code repository to image registry and then deployment processes -- using the STRIDE (\textbf{S}poofing, \textbf{T}ampering, \textbf{R}epudiation, \textbf{I}nformation disclosure, \textbf{D}enial of service, and \textbf{E}levation of privilege) threat modeling framework. 
%In particular, we use the STRIDE~\cite{lin2009a} (\textbf{S}poofing, \textbf{T}ampering, \textbf{R}epudiation, \textbf{I}nformation disclosure, \textbf{D}enial of service, and \textbf{E}levation of privilege) threat modeling framework to study the threat landscape of the containers supply chain process -- spanning code repository to image registry and then deployment processes. 
%STRIDE is proposed as it is one of the most mature threat modeling method with moderately low false positives rate or having a low number of incorrect threats, and is being implemented as part of the Microsoft Secure Development Lifecycle \cite{shevchenko2018a} \cite{scandariato2015a}. 
We choose STRIDE %is proposed 
as it is one of the most mature threat modeling framework, which has also been %is being implemented 
widely used in the %as part of the 
Microsoft Secure Development Lifecycle \cite{shevchenko2018a} \cite{scandariato2015a}. 
%In addition, STRIDE has a moderately low rate of false positives or a small average number of incorrect threats, which means that it has decent "correctness" in threat analysis \cite{scandariato2015a}. It is structured with the defender's view in mind and facilitates the development of mitigation strategies \cite{shevchenko2018b}. 
%Finally, 
STRIDE has also been successfully %applied in
adopted by several research works \cite{khan2017a}, \cite{sion2018a}, \cite{mead_2018a}, \cite{arahasanovic2017a}.
%} 
%\todo{MO: What does it mean for a threat methodology to have low false positive rate?[Way: added more details in above para]} 
Using STRIDE, we first design a data flow diagram (DFD) of the container system to map %the container system 
its %constructs 
components and their relationship via the flow of data. We then conduct a wide-range of security analysis on each component %in the system 
to discover the vulnerabilities, the associated threat actions and the resulting consequences. 

After completing our threat modeling, we then conduct a comprehensive survey on the vulnerabilities and security threats identified. % by our threat modeling. 
In particular, we analyze and discuss the effectiveness and limitations of existing mitigation strategies designed against the vulnerabilities and threats identified through our threat modeling.   
%
%Finally, 
Furthermore, we highlight open security problems and future research directions in containers security, % in containers security 
which may %ignite 
motivate the community to carry out further research in this area.

%\hl{[Way: Hi Martin, thanks for the comments. Let us know if the above edits are sufficient]} 
%
%\todo{MO: Thanks, goes in the right direction. I think we should stress more the connection of STRIDE with the survey nature of the paper, see my comments at the beginning of Sect 3 and 4, but we should do so also in intro, perhaps abstract and conclusions. The way I see it, the threat modeling was useful to identify existing research directions and match them to the literature, but also future research directions and research gaps. \\Eyasu: Thank you Martin for this important comment. Yes, we missed the connection between the survey and threat modeling in the write-up. More specifically, we did not discuss our motivation and purpose of doing the threat modeling for the survey paper. Our motivation to do the threat modeling (using the STRIDE framework) is to identify the vulnerabilities and threats in the container ecosystem first so that our survey would be based upon the already identified threats \& vulnerabilities. In other words, the threat modeling helps to specify the scope of our survey. However, as you noticed, the existing writ-up does not seem to reflect as such. So, we will update the relevant sections accordingly.}\\
%
%\indent Finally, we provide a comprehensive report, which would equip researchers and practitioners with a state-of-the-art information on the security of container systems. %landscape of containers.
\indent In sum, we believe that this work would provide a comprehensive and state-of-the-art information to researchers and practitioners on the latest security landscape of container systems. This can help the community to better understand the latest security %landscape of 
issues in containers and the available mitigation strategies to counter them.\\
\indent \emph{Organization}: The rest of the paper is structured as follows: Section \ref{background} provides relevant background information on containers, STRIDE framework, and related works on containers security. Section \ref{threat_modeling} discusses our threat modeling of the container ecosystem using the STRIDE framework. Section \ref{mitigations} investigates the existing mitigation solutions and analyses their limitations. In Section \ref{summary}, we summarize the results of our survey, %, raise open security problems, 
and highlight future research directions.
%Section \ref{future_irections} discusses open problems in containers security and highlights future research directions. 
Finally, Section \ref{conclusion} concludes our paper. %\hl{[Way: thanks and nice!]} 
\section{Background}\label{background}

\subsection{Overview of Containers}

A container is an independent, self-sufficient package for running an application or service. It includes the application binaries, the software libraries or dependencies, and the hardware requirements needed to run it, all combined into a self-contained unit. The key capabilities which enable a container to perform its function securely and efficiently (i.e., without resource constraints) are namespaces and control groups (cgroups). Namespaces provide process isolation and enable multiple application processes in containers to share a single host instance. On the other hand, cgroups allocate the host resources, such as CPU and memory, among the processes \cite{ibm2021a}. \\
\indent Containers are receiving high popularity and being widely adopted by various enterprises. This is mainly because of the following reasons: (a) a container is more lightweight than a virtual machine and therefore starts and stops much faster; (b) a container is portable as it includes the application and all its dependencies, libraries and binaries packaged into a runtime environment, therefore allowing it to run anywhere from a desktop to a datacentre; (c) a containerized application is scalable and can easily add or reduce the number of containers to meet varying demands.\\
\indent The industry's main use of containers are often tied to microservices and the cloud. Containerization supports the microservices architecture very well \cite{golden2021a}. Microservices structure an application into a set of loosely coupled software services that run in containers \cite{liu2020b}. The entire container platform and the microservice architecture are typically deployed in the cloud infrastructure as it is scalable and resilient. IBM forecasts that within the next two years, 59\% of all enterprise applications will be developed with microservices \cite{ibmcloud2021a}, further spurring the growth of container usage.  There are many enterprise-level implementations of microservices on containers, and some prominent examples are Amazon, Netflix, The Guardian, Twitter, PayPal, Tencent, Baidu, Taobao, etc. \cite{liu2020b}.
\subsection{Overview of STRIDE}\label{stride}
STRIDE is a threat modeling framework developed by Microsoft to be used by its developers during the software development life cycle. More specifically, it is used to identify and analyze vulnerabilities and threats with respect to the authentication, authorization, confidentiality, integrity, non-repudiation and availability security properties. %\sout{aspects of a system or data security.}
The STRIDE threat modeling can be performed using the STRIDE-per-element or STRIDE-per-interaction approaches \cite{khan2017a}. The former is used to analyze threats on system components, and the latter is used to analyse threats on the interaction between a pair of components. %In this paper, we adopt the former approach as we aim to analyse the security risks of each component in the container ecosystem. 
%\sout{the behaviour of each system component for potential security threats.} 
%
% \sout{It is shown that the STRIDE-per-element approach yields better results in terms of productivity, precision and recall compared to the STRIDE-per-interaction approach which analyzes the interaction between a pair of components} \cite{tuma2018a}.\\
% \comm{Comment 2: Is our analysis only on system components? I guess the communication (i.e., interaction) between different components (e.g., Github and Docker Hub) is included in the threat analysis.}\\
%
% \todo{To re-arrange these few paras}\\
% \sout{
% STRIDE-based threat modeling will be performed for a common use-case of a developer developing his application code and uploading it to a code repository which in this case is GitHub. It is one of the most popular repositories with more than 100 million source code repositories used by more than 56 million developers from more than 3 million organizations} \cite{github-a}. \sout{GitHub is used for continuous integration where a team of developers will work on different parts of the codes and later merge them into the main branch. He will then build the app image from the source code in GitHub and push it to the Docker repositories in the Docker Hub registry. The image is finally pulled to a Docker Host and to run as a container application.}\\
%
A STRIDE threat modeling %and security analysis 
is performed using the data flow diagram (DFD) of the system. A DFD is a visual representation to show the flow of information or data through a process or system \cite{vonscheel2015a}.
%\sout{and it is used for creating an overview of the system and to show the kind of input and output from the system .}
%\todo{Add a high-level description of DFD (just define DFD in one sentence).}
%The DFD 
It uses four symbols to represent system components and their relationship with others: (a) external entity such as the developer, endpoints, attacker, (b) process such as the application, a functionality, (c) data flow, which is the communication data, and (d) data store such as database, logs, and files \cite{khan2017a}.\\
\indent In general, %The STRIDE 
threat modeling using the STRIDE framework involves the following main steps: (a) drawing the DFD of the system; (b) identifying vulnerabilities on each DFD component; (c) analyzing potential threats that exploit the vulnerabilities; (d) proposing mitigation strategies for the vulnerabilities and threats identified \cite{khan2017a}.

\subsection{Literature Review on Security of Containers} 
As highlighted in the introduction, containers are vulnerable to a wide-range of cyber threats. The threats may target various attack surfaces in containers and their subsystems. The main attack surfaces of containers are user credentials, application codes, container images, container privileges, repositories, and network channels \cite{gamage2019a}. For example, stolen user credentials at the GitHub and Docker Hub can lead to user’s account being hijacked or spoofed, resulting in malicious codes and images to be uploaded into these registries. 
An attacker may also use a compromised container as a backdoor to do illegal activities on other containers. This means that if the attacker gets access to the compromised container, it can penetrate to the host kernel and launch other containers for illegitimate purposes, e.g., crypto-mining \cite{cimpanu2019b}.\\
%\footnote{https://www.zdnet.com/article/a-hacking-group-is-hijacking-docker-systems-with-exposed-api-endpoints}.
\indent The application code is another attack surface where bad coding practices can result in vulnerabilities like SQL injection, cross-site scripting, and server-side request forgery, among others. 
The Docker Hub is a popular registry for about four million of images and there are almost half which contain malware \cite{barua2020a}. Some malicious images can stay online in Docker Hub for a year and while some have been installed for more than a million times \cite{cimpanu2018a}. Therefore, if a developer creates a multi-stage Dockerfile and uses multiple images without proper scanning, he may create a container with embedded vulnerabilities. An attacker can then gain access to a compromised container and raises its privilege to gain root access to the host kernel. Lastly, there are network-related threats in the virtual ethernet bridge connected between the containers and from the internet into the container.\\
\indent %In this section, we discuss %our review of 
There are a wide-range of related works on containers security. Below, we discuss the most relevant ones. To simplify our discussion, we categorize them as ``vulnerability analysis'' and ``mitigation strategies''. 

\subsubsection{Vulnerability analysis}
There are several %researches which we list here 
existing works focusing mainly on the investigation and analysis of threats and vulnerabilities around the container ecosystem. One research initiative~\cite{lin2018a} gathered 223 container related exploits from a public database\footnote{https://www.exploit-db.com/} and classified them into a two-dimensional attack taxonomy. One dimension was the hierarchical layers of web app, server, library and kernel, and the other dimension was the consequences of attacks, %which are 
such as sensitive information leakage, remote control, denial of service, and kernel privilege escalation. However, the main emphasis of this work was on the privilege escalation exploits and how to configure the kernel security mechanisms to defend against them.\\
\indent Another study~\cite{combe2016a} %focuses on the 
was conducted on attacks %vectors of 
that mainly target the Docker platform and the image distribution process. The study %found 
revealed that insecure configurations and weak access controls of the %container 
Docker platform can lead to unauthorised access to the host filesystems and network stack of the container. Automated builds and the use of webhooks during image distribution was shown to allow a tampered code to be deployed in a production server within minutes.
% \sout{However, the focus on the vulnerabilities on Docker usages limited a comprehensive study of the threats facing the container ecosystem from image creation to image distribution.}
However, this study only focuses on threats to the Docker platform in containers. Therefore, it is not comprehensive enough to cover the multifaceted threats facing the container ecosystem from image creation to image distribution.
\\
\indent MITRE recently released the Adversarial Tactics, Techniques, and Common Knowledge (ATT\&CK) for containers. It categorized the attacks techniques on containers and the orchestration manager (Kubernetes) under initial access, execution, persistence, privilege escalation, defense evasion, credential access, discovery and impact \cite{burns2021a}, \cite{mitre2021b}. However, the MITRE framework %is focused 
only focuses on the adversary techniques and does not trace the use-case of containers nor recommends context-relevant mitigation actions.\\
\indent %The authors in research 
Another research was conducted on the security of the Docker platform by analyzing the vulnerabilities listed in Common Vulnerabilities and Exposure (CVE) \cite{duarte2018a}. In this work, %Duarte and Antunes
the authors used static code analysis (SCA) tools on the vulnerable and patched versions of the Docker's code-base in order to study the differences between the two and the effectiveness of SCA tools in detecting the vulnerabilities. This study primarily used static code analysis tools to analyze Docker's source-code and did not relate them to real use-cases nor recommend practical mitigation plans. A survey by Sultan et al. \cite{sultan2019a} %of the security issues 
was also conducted on the security of containers based on a four-dimensional risk analysis: %on four use-cases of container and they are 
risks from the application in the container, risks from another container, risks from a container to the host, and risks from the host to the container. 

Wist %in research 
et al. \cite{wist2020a} scanned 2,500 Docker Hub images, mapped their vulnerabilities using the Common Vulnerability Scoring System (CVSS), and compared the vulnerabilities across the types of images, the types of scripting languages, and packages. %Finally, the authors in 
%\sout{Another research} \cite{flauzac2020a} \sout{also compared and reported the security of different container solutions of LXC/LXD, Singularity, Docker runc, Kata-containers, and gVisor with respect to the isolation of system resources such as storage, network, processor, and memory.}
In another research, Flauzac et al. \cite{flauzac2020a} reviewed the native containers security by %performed 
conducting a static comparison of 6 container runtime solutions, namely LXC (Linux Containers), LXD (an open-source container management extension for LXC), Singularity, Docker (runc), Kata-containers (kata-runtime) and gVisor(runsc), in terms of their abilities to isolate system resources such as storage, network, processor, and memory. However, this is carried out in the container's default and standalone state and therefore does not reflect a real operating environment that is used by a container.

\subsubsection{Mitigation strategies}
While there are several %attack surfaces 
vulnerabilities and threats in the container ecosystem, %which are susceptible to threats, 
there are also certain mitigation strategies developed against them. Some of the mitigation strategies, e.g., \emph{namespaces} and \emph{cgroups}, are built-in %solutions %already built in 
to the container’s host operating system. The container’s namespaces isolate the resources of inter-process communication (IPC), mount (or filesystems), process identifier (PID), network, user (User and Group IDs), and UTS (hostnames and domain names). The cgroups control the amount of resources (like the CPU, memory, disk I\slash O) a container can use so that other co-resident containers can obtain their fair share of the resources \cite{grattafiori2016a}.\\ 
%In addition, the container is supported by 
\indent The other mitigation strategies are the underlying security features of the host kernel. These include \emph{capabilities}, \emph{secure computing mode (seccomp)}, \emph{security-enhanced Linux (SELinux)} and \emph{AppArmor} \cite{grattafiori2016a}.  The ``capabilities'' %is a 
are list of privileges that can be enabled or disabled for a process, and they serve to limit a root-enabled process from getting more than the minimum permissions required for it to perform its function. The secure computing mode (seccomp) helps to filter the system calls to the kernel from the container \cite{lei2017a}. It provides a finer control than capabilities and restricts the number of system calls an attacker %can use 
may perform %on the kernel from the container
from the container to the kernel \cite{redhat2021a}. SELinux is integrated in Centos/RHEL/Fedora distros, and it provides mandatory access control (MAC) policy setting for the applications, processes, and files in a container such that it can prevent root-enabled process within a container to illegitimately access objects outside. AppArmor is integrated in Debian/Ubuntu distros, and it is an alternative MAC %access control 
to SELinux. %While SELinux applies labels to files and subject the labels to access control rules, AppArmor applies the rules directly with the paths of the files.
While SELinux applies security rules on files, AppArmor applies the rules on file paths.\\
\indent However, recent studies revealed that most of the existing mitigation strategies of containers have certain flaws and limitations. For example, %recent studies revealed that 
the Linux-based mitigation strategies used in containers, such as cgroups, namespaces and capabilities, are subjected to attacks resulting in resources exploitations, denials of services, and privilege escalation \cite{gao2019a}, \cite{martin2018a}, \cite{stoler2021a}. 
%\sout{There is a study which shows that a container is secure with the above security features and if it is run in “non-privileged” mode and adding firewall filtering (like ebtables) to the networking connectivity bridge} \cite{bui2015a}. 
%\revised{A study from \cite{bui2015a} shows that a container can achieve effective security with the use of: namespaces, cgroups, network filter such as ebtables, MAC's measures such as SELinux or AppArmor, and running in a "non-privileged" mode.} 
Thanh Bui~\cite{bui2015a} discovered that a container cannot achieve effective security by using only the built-in security features of the host operating system, such as namespaces and cgroups. But, it should also use firewall rules (e.g., ebtables), MAC measures (e.g., SELinux or AppArmor), and run in a "non-privileged" mode. A detailed discussion of other existing mitigation strategies is also provided in Section \ref{mitigations}.\\
%\hl{Therefore, we believe there is no known work which survey the threat landscape using STRIDE on the supply chain of container from the app code to container image and finally to container deployment.}
%
%
\indent In general, most of the existing works (both in vulnerability analysis and mitigation strategies) focus only on certain security issues, and do not provide a comprehensive security analysis on the overall container ecosystem. As discussed above, some of the existing mitigation strategies have also their own limitations. Some of the %existing studies are also quite old, 
related works are also likely outdated, and they might not show the current threat landscape in containers. Therefore, it would be difficult to get a comprehensive and state-of-the-art information on the security landscape of containers. In this work, we perform threat modeling and a systematic survey %framework to do a systematic and security analysis on containers, 
on the security of containers, covering vulnerabilities, threats, threat consequences and existing mitigation strategies, to provide a comprehensive and latest information on the threat and security landscape 
of containers. %In particular, we use the STRIDE modeling framework to study the threat landscape of the containers supply chain process -- spanning code repository to image registry and then deployment. Furthermore, we assess whether the existing mitigation strategies are sufficient to ensure the security of containers. 
% ============================================================================================================================================

\section{Threat Modeling using STRIDE}\label{threat_modeling}

This section discusses our STRIDE threat modeling for containers. As highlighted in the preceding sections, we first perform threat modeling using the STRIDE framework, particularly using the STRIDE-per-element approach, to identify potential vulnerabilities and threats that may exist in each component of the container ecosystem. The main purposes of doing the threat modeling are to specify the scope of our survey based on the threats identified, map existing literature to those threats and highlight missing research angles.

%In this work, we adopt the STRIDE-per-element approach \sout{as we aim to analyse the security risks of each component in the container ecosystem} \revised{with the aim to map existing literature to potential threats which can occur in each element, and in the process highlight missing research angles}.

%\todo{MO: Maybe could say as a motivation that the purpose of doing the threat modeling with STRIDE is to map existing literature to potential threats and highlight missing research angles}

\subsection{Plotting the DFD of Containers}\label{dfd} 

\begin{figure*}[htb]
   \centering
   \includegraphics[scale=0.7]{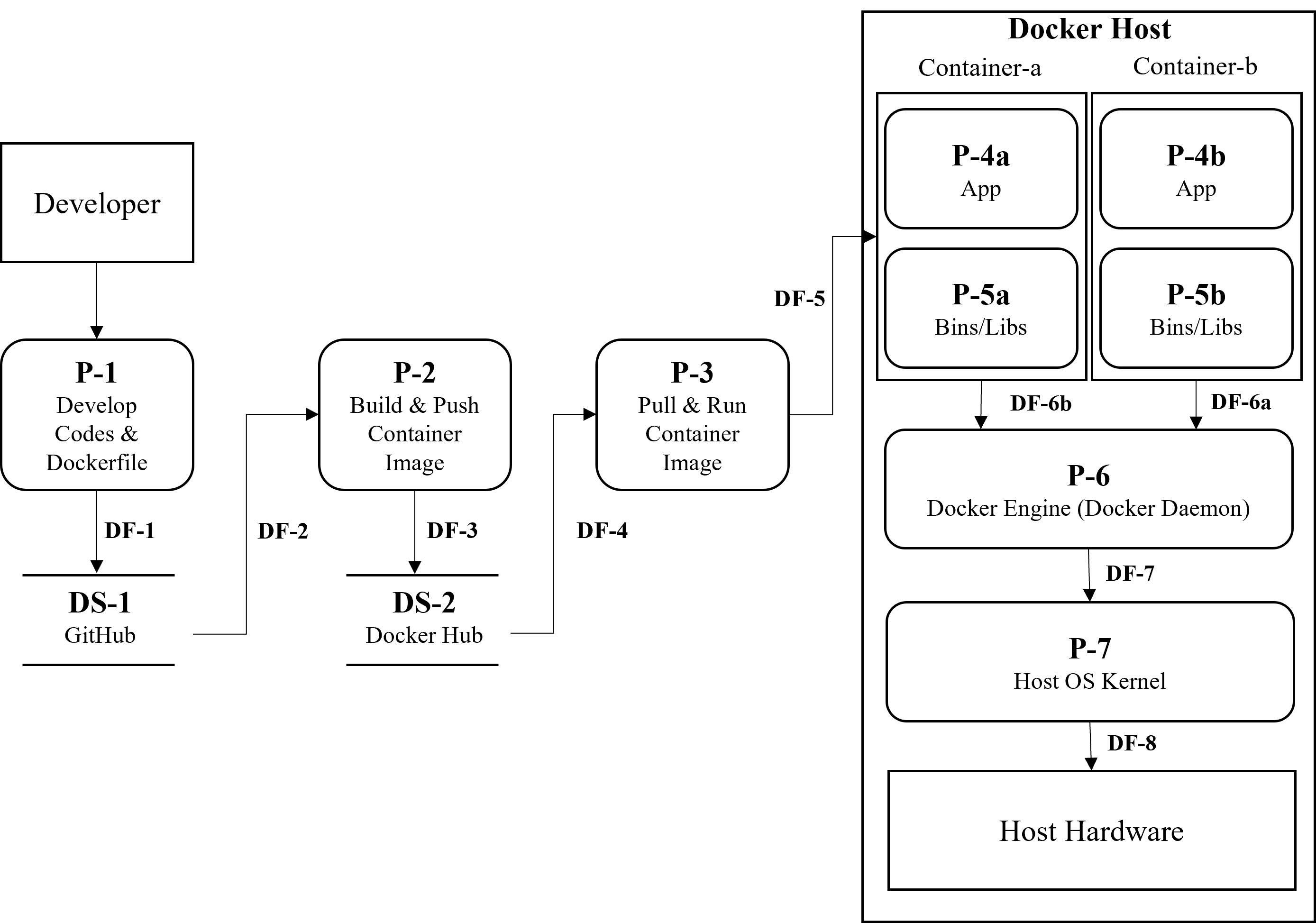}
   \caption{Data flow diagram of the container system}
   \label{fig1}
\end{figure*}

\indent As discussed in Section \ref{stride}, plotting the DFD of the system is %a crucial 
the first step in the STRIDE threat modeling. In the context of containers, a common use-case is that the developer develops his application code and upload it to a code repository, such as GitHub \cite{cito2017a}.
%
%Github is one of the most popular repositories with more than 100 million source code repositories used by more than 56 million developers from more than 3 million organizations \cite{github-a}. Github is used for continuous integration where a team of developers will work on different parts of the codes and later merge them into the main branch. 
He will then build the app image from the source-code in GitHub by creating the Dockerfile \cite{cito2017a} and pushes it to the Docker repositories in the Docker Hub registry. The image is finally pulled to a Docker Host and deployed as a container application. We plot the DFD of the %container systems 
above process in Figure \ref{fig1}, illustrating the container creation and deployment processes and its system components. More specifically, the developer (an external entity) performs the process of coding and Dockerfile creation (P-1). Then, the completed code and Dockerfile is committed and uploaded (DF-1) to the code repository GitHub (DS-1). Thereafter, the code and its libraries are packaged into a docker image (P-2) which will be pushed (DF-3) to the Docker Hub registry (DS-2). The docker image will then be subsequently pulled and run (P-3) via DF-5 into the Docker Host and deployed in container.\\
%\sout{Vulnerabilities can occur in the image during creation, in its push and pull connections, verification, during the registry storage process, and even during the communications between containers or with the host kernel.} \comm{Moved to Section \ref{vulns} below.}
%
\indent In our example, the Docker Host comprises of 4 functional components with two containers (a and b).
% \sout{Study has shown that a median Docker organization runs eight containers in a host} \cite{datadog2018a}.
The container is a wrapped and controlled environment and contains the application component (P-4a and P-4b) and the dependent libraries and binaries component (P-5a and P-5b). The Docker engine or daemon component (P-6) is responsible for launching the containers and to control their isolation level, capabilities restrictions and security profiles. The host OS kernel component (P-7) manages functions such as memory, files system, network, and process management. The Docker engine communicates with the host OS kernel using system calls. %\sout{Vulnerabilities can occur in the misconfiguration of the Docker Host and the Linux kernel.} \comm{Moved to Section \ref{vulns} below.}
% ===============================================================================================================================================
\subsection{Identifying Vulnerabilities in Containers} \label{vulns}
Vulnerabilities are the weaknesses in a system that allow an attacker to gain access into it via malicious techniques. In containers, vulnerabilities can occur during image creation, in its push and pull connections, verification, during the registry storage process, communications between the container and the OS kernel, and during the communications between two different containers. Vulnerabilities can also occur because of misconfigurations of the Docker Host and the Linux kernel.\\
\indent Using the STRIDE framework, we discovered %the following
several vulnerabilities on the DFD (cf. Fig. \ref{fig1}) of the container systems. To save space and simplify our presentation, we only discuss the most relevant ones, as shown below.
\begin{enumerate}
    \item V1: Docker Hub does not enforce stringent password policies other than the minimum password length restriction of 9 characters \cite{docker2021k}. GitHub mandates an account password to be at least 8 characters long if it includes a number and a lowercase letter, or a 15 characters with any combination of characters \cite{github2021b}. Both Docker Hub and GitHub also do not enforce the additional protection of multi-factor authentication. Therefore, a determined attacker can deploy a variety of password attack techniques like brute-force, dictionary, password spraying, and many others \cite{sailpoint2021a} to steal account IDs and passwords.
    \item V2: Docker Hub allows a developer to upload (or push) an image that is not signed. This allows an image to be downloaded (or pulled) without validating its authenticity \cite{docker2021l}. %\sout{and thus risks exposing spoofed and tampered codes to the deployed container.} 
    %If an image is tampered with, it can successfully be stored in Docker Hub and be used by unsuspecting developers for deployment. 
    This means that even tampered images can also be successfully stored in Docker Hub and used for deployment by unsuspecting developers.
    \item V3: Both Docker Hub and GitHub store software images and codes as they are, and they do not scan them for sensitive parameters, such as hard-coded passwords, access keys and other credentials.
    Inexperienced developers may include %these 
    such sensitive %parameters 
    information within the images and codes. On the other hand, industry practitioners have developed open-source tools, e.g., Docker Images Explorer\footnote{https://github.com/matiassequeira/docker\_explorer} and Whispers\footnote{https://github.com/Skyscanner/whispers}, to scan repositories and registries for passwords, API tokens, access keys, hashed credentials and others \cite{sequeira2020a}. Hence, attackers may use these tools to discover exposed credentials.
    \item V4: Docker images are not always safe and patched for use %\sout{and registry like Docker Hub does not monitor and track} 
    and Docker Hub does not check if the latest patches are applied. %\comm{[revised]}. \sout{the date of patch of the images.} \sout{An industry} 
    %A cybersecurity firm studied around 
    A recent study \cite{barua2020a} was conducted on 4 million Docker Hub images and discovered that 51\% of %the images 
    them had at least one critical vulnerability.
    Among them, about 6,400 were classified as malicious, of which 44\% were related to cryptocurrency mining, 23\% were due to flatmap-stream malware, and 20\% were a variety of hacking tools. % \cite{barua2020a}. 
    Another study of more than 2 million images from Docker Hub found that it took 181 days on average for a software originator to fix a software vulnerability, but it took an extra 422 days on average for the developer to patch the fix in the image containing the software \cite{liu2020a}. Therefore, a software with security vulnerabilities can remain in an image for more than 600 days on average and has a high probability to be downloaded and potentially exploited by the attackers.
    %\item V5: An image in the Docker Hub need not be signed when stored and need not be verified when pulled during deployment. This increases the chance of a spoofed malicious image be deployed and used by unsuspecting developers.
    \item V5: The distribution of images from Docker Hub requires only the HTTP API \cite{docker2021f}. This %can potentially 
    could allow an attacker to carry out a man-in-the-middle (MITM) attack. %CVE-2017-18641 
    In fact, a recent CVE report\footnote{https://cve.report/CVE-2017-18641} revealed that a critical vulnerability was detected 
    on LXC (%which is 
    i.e., the Linux container namespace isolation technology used by Docker) that allowed a code to be download over cleartext HTTP and to omit digital-signature checks \cite{cve2020a}. This vulnerability would allow a man-in-the-middle attacker to install malicious code into the container that will run as root.%\hl{[Way: revised] [Eyasu: Good.]}
    \item V6: Container allows API endpoints to be publicly accessible on the internet, without any firewall or password protection. This can allow attackers to successfully scan the exposed APIs and access the containers to launch attacks \cite{docker2021n}.
    \item V7: According to \cite{docker2021h}, 44\% of developers use Continuous Integration/Continuous Delivery (CI/CD) process to deploy containers. %\sout{The CI/CD pipeline consists of automated integration process that pushes the application code through the commit, build  and test phases to the code repository and subsequently to the image registry, and the automated deployment process deploys the application in a container with environment-specific parameters.} 
    The continuous integration stage pushes the application code through the commit, build  and test phases to the code repository and subsequently to the image registry. The continuous delivery stage then deploys the application in a container with environment-specific parameters.
    The entire CI/CD process presents wider attack vectors for attackers to exploit. While the automatic CI/CD process yields efficiency, the speed and lack of manual oversight creates security risks. A successful exploit in any part of the pipeline will allow an attacker to permeate its control to the rest of the pipeline.
    \item V8: %\sout{Container image is immutable and when it is built, it cannot be changed or patched, and therefore an image will frequently contain outdated packages/libraries/images.} 
    A container is immutable and when it is deployed and run, it cannot be changed or patched.
    A developer will need to ensure that the base image, application binaries and libraries are regularly updated %and 
    to rebuild and redeploy the whole image.
    \item V9: Containers are typically stateless and not appropriate to store persistent data, %and therefore 
    hence the logs %which 
    that record the containers' activities are stored in the local disk in the Docker host and in JSON file format. Each JSON log file contains only one container information \cite{sematext2021a}. Over time and as more logs are created, %the local disk will be filled and faces exhaustion unless old logs are purged or log rotation is performed 
    unless the old logs are cleared or log rotation is performed, the local disk will fill up and face exhaustion \cite{docker2021g}.
    \item V10: %One characteristic of container is its ability to connect directly with the host kernel 
    One feature of the container is that it can directly connect with the host kernel, unlike a virtual machine (VM) which requires an application to bypass the VM kernel and hypervisor. Consequently, it is easier for an attacker to access the host kernel if it can breach into an application within a container that resides on the host \cite{chelladhurai2016a}.
    \item V11: Container is reliance on Linux kernel and there are many vulnerabilities that are related to the Linux kernel that may affect the security of container, such as the vulnerability in runc module\footnote{https://www.cvedetails.com/cve/CVE-2019-5736/} that  allows a malicious container to gain root-level access to the host machine \cite{redhat2020a}. To date, there are close to 3,000 Linux CVE vulnerabilities listed by MITRE \cite{cve2021a}. However, there has not been much in-depth %studies 
    research done on the number and types of Linux vulnerabilities that directly impact containers.
    \item V12: The efficient architecture design of multiple containers on a host and sharing its CPU, memory, network, UIDs and other resources from the same kernel is also a security risk and a vulnerability. This is because, if the kernel is attacked, malicious attackers can gain root privilege of the host and from there, they can attack other containers and the entire system \cite{jian2017a}.
\end{enumerate}
%========================================================================================================
\subsection{Analyzing Threats in Containers}
Before we perform the threat analysis, %it is essential to list 
we first outline the possible threat consequences %which result from the attacker's actions. 
as we will refer them in the threat analysis sections below. A threat consequence is a security violation that happens as a result of an attack. This %can 
includes unauthorized disclosure, deception, disruption and usurpation \cite{shirey2007a}. "Unauthorized disclosure" is when an unauthorized entity gains access to the data. "Deception" is when the victim believes that a false data is true. "Disruption" is when an normal operation is disrupted and cannot carry on. "Usurpation" is when an unauthorized entity takes control of the system and operation. For %easy reference, 
simplicity, we assign %codes 
short notations for the threat consequences as follows: TC-1 for "unauthorized disclosure", TC-2 for "deception", TC-3 for "disruption" and TC-4 for "usurpation".
%========================================================================================================
\subsubsection{Spoofing} \label{spoofing}

Spoofing identity is an attack in which the attacker impersonates the victim (which can be a user, file, process, or role) to gain access into a system without the rightful consent. This attack compromises the authenticity security property, and the threat consequence is primarily TC-2. % of deception.
%In containers, spoofing attack can be made on the following accounts or 
Below, we discuss a list of potential spoofing threats in the containers ecosystem.

%\subsubsection{Spoofing of the user’s GitHub account}
\noindent \textbf{\textit{Spoofing the user’s GitHub account:}}
By exploiting vulnerability V1 listed in section \ref{vulns}, the attacker can gain access to a developer’s credential in the GitHub repo at DS-1 and to embed malware into the code. Some techniques to "steal" credentials are through spearphishing email, password-spraying, brute force, scraping published credentials in repositories \cite{sirkin2019a}, \cite{sandvik2014a}. Applying the automated deployment pipeline, the malicious code will be built into a container image at P-2. The image is then pushed into Docker Hub at DS-2 and automatically pulled and deployed at P-3 as container into the user’s docker host. The entire process can take place within minutes and may infect many other machines \cite{martin2018a}. %This is especially so if the malicious image is in the public repository instead of the private repository within Docker Hub.
The threat consequences are TC-2 %of deception
followed by TC-1.% of unauthorized disclosure.

%\subsubsection{Spoofing of the GitHub or Docker Hub}
%\paragraph{Spoofing of the GitHub or Docker Hub}
\noindent \textbf{\textit{Spoofing the GitHub or Docker Hub:}}
The GitHub repository can be spoofed by an attacker and may mislead the victim to upload his code to the attacker’s repository. The attacker can then add malicious elements into the code and upload it to the real GitHub repository. The threat consequence is %that of deception (
TC-2. The techniques can be in the form of DNS server spoofing where the attacker diverts the victim's traffic to a malicious IP address \cite{ramesh2010a} and this is achieved by using DNS cache poisoning, Kaminsky attack, or DNS hijacking (DNSpionage) \cite{kim2020a}. 
The same spoofing technique can be used on the Docker Hub (DS-2) and can lead to a malicious image being pulled to the Docker Host. So far, we have not found any article that reports about this attack vector in GitHub or Docker Hub.

%\subsubsection{Spoofing of the Docker Account and Image}
\noindent \textbf{\textit{Spoofing the Docker Account:}}
A Docker account in Docker Hub at DS-2 can be spoofed by an attacker and lead developers to go to a “fake” account to download a malicious image. The investigation team from security firm Aqua Security found that a cybercrime group created an account called “portaienr” in order to masquerade a legitimate account called “pontainer” %\footnote{https://hub.docker.com/r/portainer/portainer}
\cite{morag2020a}. %\hl{[No need to use both the footnote and citation. If the cited paper is enough, then remove the footnote. Please also lets reduce the number of footnotes in the whole paper; not good to have many footnotes.]} 
The idea was to exploit typosquatting when a victim mistyped the account name and be transferred to the attacker’s account to pull malicious images \cite{morag2020a}, resulting in the threat consequence of TC-2. %of deception.
Due to vulnerability V5, a Docker image is not scanned for vulnerability nor verified for legitimacy, hence the attack can be successful.

\noindent \textbf{\textit{Spoofing the Docker Image:}}
A Docker image can be spoofed by an attacker and lead to an incorrect image being pulled to the Docker Host. Security firm Trend Micro discovered that attackers uploaded two malicious images and labelled them as “alpine” and “alpine2” %to trick unsuspecting developers to mistake them as the popular Alpine Linux
to fake it with the popular Alpine Linux and trick unsuspecting developers \cite{remillano2020a}. Due to vulnerability V5, the image was successfully pulled without scanning. %The consequence of 
Running these images resulted in spawning of containers that installed the XMRIG crypto-mining applications. The attackers could tap on the victim’s computing resources to mine crypto-currency \cite{remillano2020a}, resulting in threat consequence TC-4.% of usurpation.

%\subsubsection{Spoofing of DNS responses to a cluster of containers}
\noindent \textbf{\textit{Spoofing the DNS responses to a cluster of containers:}}
Most application containers are deployed in Kubernetes clusters (RedHat`s survey shows that 88\% of customers use Kubernetes to manage the containers \cite{redhat2021b}) and %they 
reside in pods. Each pod communicates with each other via a bridge %which 
that runs in the root network namespace. This is made possible due to the default enablement of the capability NET\_RAW, which allows traffic (e.g., ICMP, ARP, DNS) to flow between containers. This is a characteristic of vulnerability V12 where multiple containers share the same host. An attacker can launch a DNS spoofing attack from a compromised container in a pod and return fake answers to DNS queries sent from a co-located victim container pod. Subsequently, the attacker can execute MITM attack on the network traffic between the containers \cite{sagi2019a}, \cite{chako2020a}, resulting in threat consequence TC-2.% of deception.

% ======================================================================================================
\subsubsection{Tampering}
Tampering is an attack in which the attacker modifies the data, memory space, or network and violates the security property of integrity. The main tampering threats in containers are discussed as below. 

%\subsubsection{Network between Docker Hub and Docker host tampered}
\noindent \textbf{\textit{Tampering the network between Docker Hub and Docker Host:}}
Due to vulnerability V5, an attacker can tamper DF-4 and DF-5 (Fig. \ref{fig1}), which are the data flow channels between Docker Hub and Docker Host. The attacker can insert his malicious images and be downloaded on the docker host. For example, an attacker can craft an image to contain a large file filled with garbage and when it is extracted, it would fill the host storage to cause a consequence of disruption (TC-3) \cite{martin2018a}. In another example, when the malicious image is extracted on the host filesystem, path traversals can allow the attacker to replace binaries on the host with binaries from the image \cite{martin2018a} causing the consequences of TC-1 %unauthorized disclosure
and TC-2.% deception. 

%\subsubsection{Network between GitHub and Docker Hub}
\noindent \textbf{\textit{Tampering the CI/CD pipeline:}}
This threat is due to vulnerability V7. CI/CD pipelining is a popular software development and deployment pattern used by many enterprises. The two distinct processes automate the entire flow of software build to deployment. It starts with code build, test and commit to the code repository (GitHub), to building container image based on the code, tags and pushes the container image to the container registry (Docker Hub), and finally to deploy the image as a container in the Docker host. Attacks to the network in each "pipeline" situated in DF-1, DF-2, DF-3, DF-4, and DF-5 can result in tampered software artifact and image. There are limited in-depth studies of the threats and attacks that can occur during the transportation of the codes and images along the pipelines in an automated CI/CD workflow. Martin et al. \cite{martin2018a} did a comprehensive study in the vulnerability analysis of container in three use-cases - %use in 
microservices architecture, %use for deploying whole 
virtual environment deployment, and cloud provider using it as container-as-a-service. Somya Garg and Satvik Garg \cite{garg2019a} described the mechanism of CI/CD using Docker and listed some common container security best practices in the use of namespaces, cgroups, and Linux capabilities. There is an opportunity for more research works around the security aspect of the entire CI/CD process. The consequences of this threat are TC-1, %of unauthorized disclosure 
and potentially TC-3 %of disruption 
if the network connection of any of the pipelines is disrupted.

%\subsubsection{Application codes at Docker Hub Tampered}
\noindent \textbf{\textit{Tampering application codes at Docker Hub:}}
The application code on DS-2 may be tampered with by attackers to include vulnerabilities. Docker Hub %has been
was attacked in such way before, %when the 
and the usernames and hashed passwords of 190,000 users were exposed \cite{kerner2019a}. The breach can result in the attacker accessing a user’s application image and tamper with its codes. If the image is not signed, the change will not be detected during download. In addition, Docker images may contain inherent vulnerabilities which the developers are not aware of until they are deployed in production environment. A study has shown that the official and community images contain an average of 180 vulnerabilities and 50\% of these images have not been updated \cite{shu2017a}. It takes an average of 181 days
to fix the vulnerability and an additional 422 days on average to update the image \cite{liu2020a}, and this presents a window for an attacker to exploit the vulnerability. This threat is attributed to vulnerabilities of V2 and V4. The consequences are TC-1 %of unauthorized disclosure 
and TC-2. % of deception.

%\subsubsection{Image Tampered during Image Build}
\noindent \textbf{\textit{Tampering image during image build:}}
Due to the vulnerabilities of V2 and V3 where an image is freely uploaded without any checks and controls, it can be tampered without being discovered. During the image build at P2, an attacker may inject malicious commands or vulnerable components into the image file. The image may continue to be signed and appear legitimate to the developer \cite{shen2020a}. The tampered image can cause the deployed container at P-4 and P-5 to perform malicious acts to the host or other containers residing in the same host causing the consequences of TC-1, TC-2 or even TC-4. Developers rely on open source libraries when developing their applications. A commercial study finds that seven in ten applications use at least one open-source library with a security flaw \cite{rashid2020a}, and that the library vulnerabilities increase by 88\% over a two year period \cite{tal2019a}. Palo Alto Networks conducted a study which found that 96\% of third-party container applications deployed in the cloud contain known vulnerabilities \cite{greig2021a}. The attack surface is further expanded if the libraries have their own dependencies on codes from other libraries. The malicious libraries in a deployed container at P-4 and P-5 will interact via the Docker daemon at P-6 to gain unauthorised access to the OS kernel. These threats will lead to the consequence of %unauthorized disclosure, deception and even usurpation 
TC-1, TC2, and even TC-4 when the host kernel is under control.
%\subsubsection{Mitigations}
%\noindent \textbf{\textit{Mitigations:}}
%
%============================================================================================
\subsubsection{Repudiation}
Repudiation is associated with an attacker claiming that something which is done is not performed by him. This attack violates the security property of non-repudiation. In the following, we discuss the main repudiation threats in container systems. 

%\subsubsection{Potential weak protections for audit data (logs)}
\noindent \textbf{\textit{Disabling logging functions:}} %Potential weak protections for audit data (logs):}}
An experienced attacker will cover his track to avoid detection and attribution. The attacker may attack the audit mechanism and attempt to delete or modify the logs %which is 
stored in element P7 in Fig. \ref{fig1}. He may disable the logging function using "Auditpol" in Windows systems or "auditctl" in Linux systems. He may delete the logs with clearlogs.exec in Windows systems and shred tools in Linux systems \cite{belding2019a}. The consequence is the disruption of logging activity.

\noindent \textbf{\textit{Modifying log data:}}
The log files in Docker can be found in  /var/lib/docker/containers directory on the host system \cite{rahic2020a} and they can be modified by the attacker. This threat is possible due to the vulnerability of V9 and V10 as the container is dependent on the Linux host for logging activities and storage. At this point we have not found any reports that describe a real attack event on Docker logs. The consequence is deception by modifying the log data.

%\subsubsection{Overwrite log disk space with junk}
\noindent \textbf{\textit{Overwriting log disk space:}}
A container utilizes the memory and storage space of the host and this vulnerability is aligned with V9 and V10. A container is enabled with the capability CAP\_AUDIT\_WRITE to record activities and events into the kernel audit log \cite{fiser2019a}. The kernel audit log is stored on the disk in the host at P-7 and the attacker container can write massive amount of junk data onto the disk and overwrite the valid logs recorded by the victim container \cite{luo2016a}. This attack can cover the tracks of a malicious action and prevent the victim from accessing valid logs to perform investigation. This threat will result in deception as the real logs are overwritten.
%\subsubsection{Mitigations}
%\noindent \textbf{\textit{Mitigations:}}
% ======================================================================================================
\subsubsection{Information Disclosure}
Information disclosure is allowing unauthorized entity to access data, information, processes or networks which he is not allowed to. This attack compromises the security property of confidentiality, and the following are a list of information disclosure threats in containers. 

%\subsubsection{Weak access control of GitHub repo}
\noindent \textbf{\textit{Weak access control of GitHub and Docker Hub:}}
Weak access control of GitHub (DS-1) and Docker Hub (DS-2) allows an attacker to access information which he is not authorised to do so.
There have been several security breaches in GitHub where identity keys and data information have been stolen. For example, developers from Starbucks expose API keys in GitHub, which can allow an attacker to access its active directory management platform \cite{ilascu2019a}. Starbucks later removed the repository and revoked the API keys. Another attacker got access into CircleCI’s user data which include their GitHub’s usernames, emails, repo URLs, branch names, organization names and repo owners \cite{joshi2019a}. This prompted CircleCI to enforce two-factor authentication (2FA) for their account holders. Another attack involved gaining access into all the Git hosting services including GitHub, GitLab, etc. to steal source-codes and demanding ransoms from the owners \cite{cimpanu2019a}. Some of the victims had admitted to using weak passwords and forgetting to remove access tokens for old apps. Recently, millions of Brazilian COVID-19 patients' personal private information (including the Brazil's President, ministers and provincial governors) were exposed when a spreadsheet which stored the login credentials of the government healthcare systems were exposed by a GitHub user \cite{cimpanu2020a}. The source codes of Nissan were leaked and exposed from a Git server when its developer secure it with its default username and password combo of admin/admin and they were easily cracked by attackers \cite{cimpanu2021a}. Mercedes Benz's smart car components source code were leaked when an outsider successfully signed up for an account in its Git web portal using a non-existent Daimler corporate email \cite{cimpanu2020b}. In addition to the easy access into GitHub account, an inexperienced developer may make a change in a source code file and unknowingly commit and upload all other files (which include sensitive ones) in the same folder into GitHub. An attacker who breaches a Github account can access these sensitive files.
%\subsubsection{Weak access control of Docker Hub}
%\noindent \textbf{\textit{Weak access control of Docker Hub:}}
The access control of Docker Hub at DS-2 can be exploited and sensitive data be exposed. In 2019, a database of 190,000 users' usernames and their hashed passwords in Docker Hub was hacked into by attackers \cite{matthews2019a}. On separate occasions, attackers managed to steal the credentials from the cloud providers and took control of the container instances which were owned by Aviva, Gemalto and Tesla and used them for crypto-currency mining \cite{redlockcsi2018a}. This threat can be attributed to vulnerability V1 which is due to a non-stringent credential and access control measures. The consequence is the unauthorized disclosure of sensitive information.

%\subsubsection{Sensitive parameters to access host data}
\noindent \textbf{\textit{Sensitive parameters to access the host:}}
The run-commands used in P-3 to run a container may contain sensitive parameters which allows an attacker that develops the container image to gain access to the user’s host and its data. These parameters are not usually detected by the security scanner as they are not malicious in nature. For example, a user may run a container command with “- -privileged” to access certificate on the host to spawn a container \cite{jarvis2020a}. The use of such "sensitive" parameter will allow the container to gain root access to the host and this can be %taken advantage of by 
exploited by an attacker \cite{liu2020a}. Another example is the use of “- -volume” and “-v src:dest” that allows a container to gain access to “src”, which is a volume in the host and as a result allows an attacker to upload data in the host to a online repository \cite{liu2020a}, resulting in %causing the 
threat consequence of TC-1. In some instances, there may be a need to configure the parameter of “- -pid=host" within a container in order to run debugging tools, like strace or gdb \cite{docker2021c}. Such configuration allows the container to share the host's PID (process ID) namespace. If an attacker gains control of the container, he will be able to view all the other processes running on the host. Armed with info of the PID, along with "owner" and path of the executable file, the attacker can conduct attack to other containers and the host \cite{liu2020a}. This threat is attributed to vulnerabilities V10 and V12 which is due to the common Linux kernel shared by multiple containers. Due to vulnerability V10, the configuration options of the Docker engine/daemon at P-6 can provide access to the host OS kernel. This can be achieved with the options of “-net=host”, “-uts=host”, “-privileged”, and additional “capabilities”. The option “-uts=host” can allocate the same UTS namespace for the container and the host which allows the container to see and change the host’s name and domain \cite{martin2018a}. The capability “-cap-add=SYS\_ADMIN” can enable a container to remount /proc and /sys sub-directories in read/write mode, and change the host’s kernel parameters \cite{martin2018a}, leading to potential threat consequences of TC-1 and TC-4. %unauthorized disclosure (TC-1) and usurpation (TC-4).
%\subsubsection{Leakage of information via covert channels between containers}

\noindent \textbf{\textit{Leakage of information between containers:}}
Containers that reside in the same Linux host and share the OS kernel (P-7) can leak information to each other via storage path mapping, port mapping, layer-2 network connection, and covert channels. % and 
This can enable an attacker of one container to gain access into another co-locating container \cite{luo2016a}. Some of the methods include exploiting the openly observed globally used memory (GUM), which an attacker %container 
can obtain visibility of the victim container's memory information \cite{luo2016a}; accessing the global variable of inode number (or index node) allows an attacker container to know the metadata of a victim container's process file \cite{luo2016a}; and an attacker container can read into the kernel message buffer (KMB) which is written into by a victim container with the CAP\_SYSLOG enabled \cite{luo2016a}. This is again due to the vulnerabilities V10 and V12 and the consequence is the leakage of unauthorized information (TC-1).

%\noindent \textbf{\textit{Other accesses to the host OS kernel from container:}}

%\subsubsection{Mitigations}
%\noindent \textbf{\textit{Mitigations:}}

% =======================================================================================================
\subsubsection {Denial of Service (DoS)}
The denial of service causes a service to be disrupted or degraded such that users cannot access the service. This attack violates the security property of availability. Most of the threats listed below are attributed to the vulnerability V10 which is the close connection between the container and the host kernel unlike a virtual machine which is separated by the VM kernel and the hypervisor. The attack involves abnormally consuming resources such as CPU, memory, storage, networks, etc. The threat consequence is mainly TC-3.  % or disruption. 
Below, we discuss the main DoS-related threats in the containers context.

%\noindent \textbf{\textit{GitHub or Docker Hub becomes inaccessible:}}
\noindent \textbf{\textit{Inaccessibility of GitHub or Docker Hub:}}
The attacker may cause GitHub (DS-1) or Docker Hub (DS-2) to be inaccessible to developers for code updates and container deployments. %At this point, there is no report that such an incident has happened. 
While the infrastructure facilities of GitHub and Docker Hub are not publicly known, it is assumed that they are highly resilient, secured and are distributed across multiple sites like the commercial cloud computing services, such as AWS, Microsoft Azure or Google Cloud. Therefore, at this point there is little evidence to show that the services from GitHub or Docker Hub have been disrupted due to attacks on their server infrastructures. An article was written that painted a scenario where a DDoS attack targeted at the control traffic between the Network Operations Center (NOC) and the data center's Heating, ventilation, and airconditioning (HVAC) could potentially result in overheating and to cause a data center outage \cite{zahid2014a}. However, in reality, there had also been data center outages that resulted from overheating due to equipment failures \cite{tung2020b}, service component failures such as the DNS outage in Azure \cite{tung2021a}, Kinesis disruption in AWS \cite{tung2020a}, and other non-attack related causes.

%\subsubsection{Exception handling by OS kernel}
%\noindent \textbf{\textit{Exception handling by OS kernel:}}
\noindent \textbf{\textit{Service disruption at host via OS kernel:}}
From Fig. \ref{fig1}, the container via the Docker engine (P6) communicates with host OS kernel (P7) via a series of system calls. By default, each container has access to the host's CPU cycles and memory without limit \cite{docker2021j}. An attack on the OS kernel will cause the disruption of services to the host's computing resources like the CPU, memory, storage, and others resulting in the threat consequence %of disruption (TC-3).
TC-3. Attacks utilizing exceptions handling, logs writing, and disk write-backs can impact CPU, disk I/Os and memory performances. %\newline
%\indent 
The Linux kernel will trigger an exception handler when exceptions such as faults (e.g., divide error) and traps (e.g., overflow) occur. When one of them happens, the kernel will send a signal to the process which generates it, and it will take steps to recover or to abort \cite{bovet2007a}. The exception will trigger the core dump kernel function to generate a core dump file which is used for debugging.
%In Ubuntu, the core dump application is Apport and it is spawned (for every exception) by a kernel thread and will consume the system resources shared with the container. The running of Apport process consume much resources which are taxed across all the CPU cores.
It is shown that when a container keeps raising exceptions (example div 0) and triggers the core dump, the host system CPU and memory performances are reduced by 95\% \cite{gao2019a}. Therefore, an attacker can create a DoS attack on a host using this exploitation and thereby impacting the performance of all containers which run on this host. %\newline
%\indent 
System logging in Linux at P-7 is typically performed by journald which is a part of systemd, an init system and system manager \cite{gheorghe2020a}. As a system service, journald not only collects system and kernel log messages, but it also collects three types of log messages in a container. They are switch user (su), add user/group, and exception \cite{gao2019a}. As journald is a system service, its resource utilization will be taxed on the host and is not controlled by the container cgroups. It is shown that the three container logging operations performed by journald can cost up to 20\% extra CPU utilization and an average of 2MB/s IO throughput \cite{gao2019a}. Therefore, this is an exploit which an attacker can use in a container to overwhelm the host resources which in turn impact the performance of the other containers causing a TC-3 consequence. % (or disruption).\newline
%\subsubsection{Disk writeback for data synchronization}
%\noindent \textbf{\textit{Disk writeback for data synchronization:}}
%\indent 
To improve performance, the Linux kernel writes data in the cache memory and later performs a disk writeback of the data into the disk at the host. However, data may be lost or corrupted when the system crashes, and one way for a user to invoke a writeback is to run a system call “sync”, which writes any data stored in the cache memory out to the disk \cite{haas2020a}.  It is shown that when a malicious container keeps calling “sync” while another victim container performs write operations, it leads to high CPU wait time due to the combination of sync and write operation. The victim I/O performances (such as sequential read \slash write and random read\slash write) are reduced to almost 1\% \cite{gao2019a}. This shows that an attacker can launch a DoS attack on the host and hence on another container by exploiting the data writeback mechanism to the disk.
%\subsubsection{System logging for container operations}
%\noindent \textbf{\textit{System logging for container operations:}}
%\subsubsection{Other accesses to the host OS kernel from container}

%\subsubsection{GitHub or Docker Hub becomes inaccessible}

%\subsubsection{Data flow to and from Docker Host may be disrupted}
%\noindent \textbf{\textit{Data flow to and from Docker Host may be disrupted:}}
%The attacker may disrupt the data flow across the network boundary where the Docker Host resides within the container orchestrator cluster. \textless There is no known study on the attack to the network path to Docker Host \textgreater
%\subsubsection{The Container datastore becomes inaccessible}
%\noindent \textbf{\textit{The Container datastore becomes inaccessible:}}
%Container uses persistent volume to store data permanently even when the container is destroyed or re-created. The persistent volume may be disrupted and the data become inaccessible. \textless There is no known study on the attack to container persistent volume. \textgreater

%\noindent \textbf{\textit{Any of the data flows becomes inaccessible:}}
\noindent \textbf{\textit{Inaccessibility of the data flows:}}
As shown in prior threats targeting the CI/CD automated integration and deployment process (due to vulnerability V7), an attack in any of the data flow connections at DF-1, DF-2, DF-3, DF-4, and DF-5 will cause disruptions to one or more of the processes of code commits, images build and upload, images download and containers deployment.  

%\subsubsection{Mitigations}
%\noindent \textbf{\textit{Mitigations:}}

% =======================================================================================================
\subsubsection{Elevation of Privilege}
Elevation of privilege increases the level of authorization of an attacker such that he can perform operations or access information which he is %prohibited 
not allowed to do so. This attack violates the authorization property of security and leads to the consequence of %usurpation (TC-4). 
TC-4. A container is vulnerable to a host take-over attack because of vulnerabilities V10, V11, and V12. This is due to its tight integration with the Linux kernel, sharing it with other containers, and inheriting vulnerabilities that frequently discovered in the Linux operating system. Below, we discuss the specific elevation of privilege threats in containers.   

%\subsubsection{Misconfiguration of Docker image}
\noindent \textbf{\textit{Run container as root:}}
At P-2 in Fig. \ref{fig1}, there are some considerations when configuring the Dockerfile to build a Docker image. By default, the Docker container runs as root %because 
since the Docker daemon needs root privileges to modify the host filesystems to run \cite{rice2021a}, unless a developer intentionally configures it otherwise. As such, an inexperience developer may pull and deploy container at P-3 in a root privilege mode. % and 
This allows an attacker to copy files from the host to the container and access them, and launch a remote command execution (RCE) attack \cite{pavisic2019a}.

%\subsubsection{Gain access and elevate privilege in host via misconfigured container}
\noindent \textbf{\textit{Gain root access via misconfigured networking:}}
A newly created container will be configured with the default bridge network at the Docker daemon networking stack at P-6. The default bridge network allows other unrelated containers or services to communicate with it remotely \cite{docker2021b}. An attacker can exploit this container and open a listening port to other containers in the same network. When it discovers an open port, it will connect to its Docker daemon and instruct it to download and run a malicious script \cite{shevchenko2020a}. The malicious script can potentially disable the security system of the host, create a root user, and download and install a malicious program such as a crypto-miner to perform crypto currency mining \cite{shevchenko2020a}.

%\subsubsection{Vulnerabilities in Docker images}
%\noindent \textbf{\textit{Vulnerabilities in Docker images:}}

%\subsubsection{Use of System Calls to gain privilege}
\noindent \textbf{\textit{Use of system calls to gain privilege:}}
During the starting and running of application containers, system calls are made from the containers to the host kernel at P-7.  It is noted that 331 system calls are allowed by default, but an experiment with a MySQL database container show that only 116 system calls are needed in the booting phase and 58 system calls are used in the running phase \cite{lei2017a}. In another experiment using the Apache web server, 47 unnecessary system calls are enabled in the container, and they are found to be vulnerable to exploitation with CVE security level of medium and above, e.g., signalstack()\footnote{https://www.cvedetails.com/cve/CVE-2009-2847/} and setsockopt()\footnote{https://www.cvedetails.com/cve/CVE-2017-6074} \cite{lei2017a}. Therefore, a high default number of system calls increases the attack surface and the unnecessary system calls can be used by malicious processes to gain elevated privilege in the host. 

%\subsubsection{Kernel privilege escalation attack}
\noindent \textbf{\textit{Kernel privilege escalation attack:}}
A study has shown that an attacker can make use of a compromised container to launch attack on the host kernel at P-7 to escalate its privilege. Exploits contained in CVE-2017-7308, CVE-2017-5123 and CVE-2016-8655 (or Exploit-DB IDs of 41994, 43127, 43029 and 40871) show that privilege escalation exploits can overcome the default security mechanisms in “Namespace”, “Cgroup”, “Capability”, “Seccomp” and “MAC” to launch a malicious shellcode in the kernel and in supervisor mode \cite{lin2018a}. This is carried out by bypassing the KASLR (Kernel Address Space Layout Randomization) to obtain the address of the critical kernel static functions, and to launch attacks like “use after free”, race condition, buffer flow etc., to enable the overwriting of the pointers of the kernel functions. The attacker then overwrites the kernel functions’ pointers to disable the CPU protections of SMEP (Supervisor Mode Execution Protection) and SMAP (Supervisor Mode Access Protection) and to point to a malicious user space function or shellcode, which invokes a kernel function commit\_creds() to apply for root credential \cite{lin2018a}. Another attack leverages the "time of check to time of use" (TOCTOU) vulnerability to gain root access to the host. This happens when a user executes a "docker cp" command to copy contents from the container to the host filesystem and the attacker adds a symlink component to the path after the resolution and before the operation. This results in resolving the symlink path component on the host as root allowing it to read and write to any path on the host \cite{fisher2019a}.
%\subsubsection{Mitigations}
%\noindent \textbf{\textit{Mitigations:}}
% =======================================================================================================

%\section{Potential Mitigation Solutions}
%\section{Mitigation Approaches}\label{mitigations}
\section{Existing Mitigation Strategies and Their Limitations}\label{mitigations}
%\sout{After identifying the threats using the STRIDE framework, we hereby discuss the} %potential mitigation approaches to counter the cyberattacks of these threats.
%\sout{main existing mitigation strategies against the threats and their limitations.}
%\revised{With the application of STRIDE, we successfully identify potential threats by mapping current works from the research community and the practising industry. We hereby discuss existing mitigation strategies against the threats and identify research areas that have not yet been explored.}  

In Section \ref{threat_modeling}, we discussed the potential threats and vulnerabilities we have identified using the STRIDE framework. This helped us to also explore the respective mitigation strategies mentioned in the literature as well as to identify research areas that have not yet been explored. Below, we will discuss the identified mitigation strategies and their limitations to address the corresponding security threats in containers.

%\todo{MO: Again it would be nice to have an explicit link to the survey nature of the paper. For instance: STRIDE help identifying potential threats and existing papers mentioning in the literature, we now as a result follow existing mitigations and  identify research areas that have not yet been explored.}

%
\subsection{Multi-Factor Authentication Systems}\label{MFA}
One of the practices to %protect an 
harden access to an account %include the use of 
is using multi-factor authentication (MFA) systems. It is found that 99.9\% of the accounts that were breached before did not use MFA, and that a basic 2FA using SMS could stop 100\% of automated attacks and 96\% of phishing attacks \cite{nahari2021a}. Docker Hub offers %two-factor authentication 
2FA using mobile phone authenticator application (e.g., Google Authenticator) or Yubico Authenticator with a Yubikey \cite{docker2021p},  while GitHub offers %two-factor authentication
2FA with applications like Authy, Duo Mobile, Google Authenticator, Microsoft Authenticator, etc. \cite{gitlab2021a}. In addition, we should enforce policies like strong passwords and regular rotation of passwords. Microsoft has listed useful password guidelines like banning common passwords, not to re-use organization passwords for non-work related purposes, enable risk-based %multi-factor authentication, 
MFA, and others \cite{microsoft2021a}. These measures can help prevent attackers from using stolen credentials to access codes and images in containers.

\noindent \textbf{\textit{Limitations:}}
While 2FA improves the security by adding a layer of authentication to the password controls, it does have several disadvantages. 2FA increases the time and cost to access the accounts and this can be significant if an organization has thousands of employees \cite{doffman2020a}. By default, 2FA uses SMS to text the verification code to a user's phone. An attacker can easily perform SMS attack on a compromised phone or the messaging center to retrieve the verification code %which 
that is not encrypted \cite{doffman2020a}. While using a mobile authenticator app is safer than 2FA with SMS, there is a report that shows attackers stealing one-time passcodes generated by Google Authenticator on a mobile phone \cite{cimpanu2020c}.   
\subsection{Image Security}
Securing container images is one of the existing mitigation strategies against threats in containers. Below, we discuss the main image security strategies applicable in container systems.

\noindent \textbf{\textit{Reducing attack surfaces:}}
It is recommended that an image be kept minimal so that the attack surfaces can be reduced. A couple of best practices in this regard include using multi-stage build feature that enables the developer to create an intermediate container with the required tools, and selectively copy the artifacts to the final image with only the minimal required binaries and dependencies \cite{docker2021i}. The other practice is to use distroless\footnote{https://github.com/GoogleContainerTools/distroless} images as they do not contain package managers, shells, and others so that the image is kept minimal \cite{iradier2021a}. 

\noindent \textbf{\textit{Signing images:}}
It is advised that a developer digitally signs his image with Docker Content Trust \cite{docker2021a} that is attached to the Notary server,which is used for validating the integrity of the images \cite{github-b}. Consequently, it is a good practice for developers to verify the authenticity of the images before pulling them by enabling Docker Content Trust \cite{docker2021a}. In addition, the developer should ensure that the hash of the image is the same at the Docker Hub %and 
as well as when it is deployed to the Docker host. Another effective mitigation method is to enable the Linux Integrity Measurement Architecture (IMA), which would validate the file signatures against pre-installed certificates and denies unauthorized file from being executed. It is shown that IMA can prevent a code that is not signed %code 
or signed with unknown key, or a modified code with an invalid signature \cite{sun2018a}.

\noindent \textbf{\textit{Vulnerability scanning:}}
After building the image and before a developer pushes the image to Docker Hub, one good practice is to scan the image by baking a scanning command in the Dockerfile or running a script at P-2. Another good practice is to scan the images before deploying them. Docker Hub provides vulnerability scanning but only to paid subscribers under the Pro or Team plan \cite{docker2021d}. However, there are several open-source container scanners in the market and these are Anchore, Clair, Dagda, OpenSCAP, Sysdig Falco, and others \cite{bhat2018a}. In addition, for static scanning, we can also perform dynamic analysis by running the container in a Docker-in-Docker sandbox mode and scanning it with tools such as VirusTotal (a collection of anti-virus tools) and examining the collected tcpdump/log files for file changes, network traffic, and list of processes \cite{brady2020a}.
For application code scanning, GitHub offers CodeQL \cite{github2021c} and integration to third-party code scanning tools, such as Checkmarx, Synopsys Intelligent Security Scan, Veracode Static Analysis, and others \cite{palafox2020a} for identifying vulnerabilities in the codes.

\noindent \textbf{\textit{Limitations:}}
While signing the image is an important safety measure, the private keys used for signing can be stolen. There have been several instances and methods deployed to steal private keys \cite{drd2020a}, \cite{appviewx2019a}, \cite{goodwin2019a}, and therefore more research can be done to protect them.
With respect to container scanners, %research 
Javed and Toor \cite{javed2021a} used Claire, Anchore, and Microscanner to investigate the quality of the container scanning and found that they were at most 65\% accurate in the detection rate, leaving about 34\% of the vulnerabilities being undetected and passed through and being deployed in production environment. The container scanners depend on the CVE data from public databases such as the National Vulnerability Database from the National Institute of Standards and Technology (NIST), Red Hat Enterprise Linux, Debian, and others to check if an image has vulnerabilities. As such, the scanners are not able to detect security flaw that has not been publicly disclosed or if the image is rebuilt from an open-source software package and given a version number which is not tracked in the vulnerability databases \cite{avner2021a}. Another limitation is the disparate processes and tools across the container scanning workflow and there is no one integrated tool which can perform static and dynamic scans.      
\subsection{Security Patching}
It is advisable for developers to use verified and official images from trusted repositories and providers. A study \cite{wist2020a} shows that "official" images are the most secure among image types, which include "verified", "certified" and "community". Both the "official" and "verified" images are the most updated, while the "community" and "certified" images are the least updated ones. The developers should update their images with the latest security patches and rebuild the images periodically. %The National Institute of Standards and Technology (NIST) 
NIST recommends the following scenarios and the urgency of patching \cite{souppaya2020a}. Routine patching is the standard procedure to patch on a regular release cycle (e.g. Patch Tuesday). Emergency patching is carried out quickly to address extreme severity vulnerabilities and exploits. Emergency workaround is performed prior to the vendor releasing a patch and it may include roll back exercises. Lastly, it can involve the isolation of unpatchable assets if the systems cannot be easily patched \cite{souppaya2020a}.\\
\indent IBM researchers Araujo and Taylor \cite{araujo2020a} developed a just-in-time (JIT) patching framework called "Insider" for patching running legacy application processes. This was done by injecting and compiling the code inside the running processes while sandboxing malicious processes for threat investigations. However, it is not developed for a containerized environment. A containerized application self-patch framework was developed by Tunde-Onadele et al. \cite{onadele2020a} that performed attack detection by using machine learning methods on the system calls; attack classification by comparing it to the CVE database; and finally patch execution by downloading the latest files to update the image and spinning new application container.
%However, there is no such patching program for containerized application and is an important area for future development.

\noindent \textbf{\textit{Limitations:}}
At this point, there is no known automatic or JIT patching mechanism developed for the container. While rapid patching is important to address vulnerability in the container before an attacker gets into it, it may cause compatibility issue with the application without first testing it in a lab environment. Therefore, a reliable and rapid patching framework for containerized application is a gap which should be tackled quickly.
\subsection{Minimise Administrative Privileges}
One way to mitigate against attacks on sensitive parameters is to design a mechanism to detect sensitive parameters and to alert the user of the risks before he executes the run command. As far as we know, there is no such mechanism available to perform this function in containers. There are recommendations by Center of Internet Security (CIS) to limit harmful docker run options and some examples are, hardening host configuration, limit file permissions, configure TLS for Docker Hub and control socket, and many others \cite{martin2018a},\cite{cis2021a}. There are also methods to configure a container to run in a "rootless mode" and some of these are proposed by Docker \cite{docker2021m}, Bitnami \cite{godoy2018a}, Redhat \cite{walsh2019a},\cite{mccarty2019a}, and others. 

System calls related vulnerabilities could lead to privilege escalation attacks. Almost 17\% privilege escalation attacks listed in the Exploit Database maintained by Offensive Security were due to system calls \cite{provelengios2018a}. The threats of mis-using of the system calls in the containers can be mitigated using the following methods. The SPEAKER mechanism developed by Lei et al. \cite{lei2017a} traces the systems calls needed in the booting and running phases of a container and then dynamically modifies the security filter to reduce the number of system calls in each phase, thereby reducing the attack surface which is exposed by the system calls. 

Another method, called Classified Distributed Learning (CDL), which is developed by Lin et al. \cite{lin2020a}, uses the machine learning algorithm to detect anomalous behaviour of the system calls and to raise an alert if it differs from the normal pattern. The system calls are collected from running containers and they are classified by application class using the random forest technique and subsequently grouped together. The autoencoder neural network is then used to train on the system calls data set and the model is applied to new system calls flow to detect anomalous behaviour \cite{lin2020a}. The accuracy rate is 74\% when applied to 24 commonly used applications with 33 known vulnerabilities. %\\

Another method of anomaly detection developed by Abed et al. \cite{abed2015a}, uses the Bag of System Calls (BoSC) technique. This method is first introduced in 2005 to improve the then widely used fixed-length contiguous subsequence models in intrusion detection systems (IDS) \cite{kang2005a}. It is subsequently applied onto the Linux containers to detect anomaly in system calls \cite{abed2015a}. The method collects “bags of system calls” (BoSC) in a normal container operation and stores them in a database. In a new container operation, the new bags of systems calls are compared against the database of BoSC and if there are mismatches which exceed a certain threshold, an anomaly is assumed. Each BoSC consists of an array of distinct system calls’ frequency of occurrences \cite{abed2015a}. The method is shown to be accurate to detect anomaly but it is only tested on a MySQL container using SQL injection attacking tool. It has not been proven to work in other use-cases.

Rastogi et al. \cite{rastogi2017a} developed a method called Cimplifier, which applies the principle of privilege separation and it aims to partition a container into smaller containers which isolate from each other and only equip with the necessary resources and they communicate with each other when needed. Lastly, it is also a good practice to limit the permissions of capabilities in the container to those which are necessary so that attackers do not take advantage to exploit them to gain control of the host \cite{owasp2021a}.

\noindent \textbf{\textit{Limitations:}}
The system calls anomaly detection techniques proposed are either not highly accurate or only tested on a specific use-case. There is a need to develop higher accuracy anomaly detection method which can apply to most use-cases and applications.
\subsection{Proper Isolation}
The cgroups of the Linux kernel are primarily functioned to control and limit the underlying host resources for each container. Within the cgroups, there is the cpuset subsystem which a developer can configure to bind a container to a set of CPU cores so that the CPU resources are protected from DoS attack \cite{chen2018a}. It was also demonstrated that the use of Linux memory bandwidth management module MemGuard can limit the CPU access to the memory and can thus prevent a DoS attack on the memory \cite{chen2018a}. There are numerous security best practices that can mitigate DoS attacks, e.g., using read-only filesystems, limiting kernel calls, restricting networking and inter-container communication, not expose Docker daemon socket, limit resoucres, and others \cite{chelladhurai2016a}, \cite{owasp2021a}.

\noindent \textbf{\textit{Limitations:}}
The use of cgroups and namespace isolation methods in containers have several limitations. A recently discovered CVE vulnerability\footnote{https://nvd.nist.gov/vuln/detail/CVE-2020-25220\#vulnCurrentDescriptionTitle} showed that a use-after-free flaw can occur in the cgroupv2 subsystem %when the 
during system reboot. This flaw would crash the system or escalate its privileges \cite{redhat2021c}. The other limitation of container isolation is that the current isolation measures do not truly sandboxed containers that share the same host \cite{chen2019a}. Consequently, numerous container escape vulnerabilities have been discovered, such as CVE2014-3519, CVE-2016-5195, CVE-2016-9962, CVE-2017-5123, and CVE-2019-5736. %Research
Gao et al. \cite{gao2019a} also presented several exploiting strategies to escape the resource protection set up by the cgroups. Furthermore, other researches \cite{martin2018a}, \cite{bui2015a} showed that the current container isolation system cannot effectively isolate the network as the same network bridge is shared by the containers, causing ARP poisoning and MAC flooding attacks on the containers.
\subsection{Prevent Confidential Data Leaks}
To mitigate against credentials exposure, it is a good practice not to store unencrypted secrets in Git repositories, but to use tool like git-secret to encrypt passwords, secret keys and sensitive data \cite{wallen2020a}. Within Docker Hub, developers can store secrets in credential stores such as D-Bus Secret, Apple macOS keychain, Microsoft Windows Credential Manager and "pass" \cite{docker2021e}. The recommendations to strengthen passwords and protect access control as described in section \ref{MFA} are applicable here.

When committing and uploading modified files into GitHub, one good practice is to use ".gitignore" feature to specifically exclude certain files from being "committed" into GitHub \cite{github2021a}. This will prevent sensitive files which reside in the same folder as the program code to be uploaded into GitHub. Another practice is to use ".gitignore" to whitelist the files (instead of exclude) to commit \cite{kuizinas2020a}.

\noindent \textbf{\textit{Limitations:}}
Credential storage secrets manager or vault is not bullet-proof. CyberArk had tested a method to steal credentials stored in Local Security Authority (LSA) Secrets registry and to achieve lateral movement throughout the system \cite{naim2016a}. Despite having solid vaults, confidential data and credentials can be leaked if the user share credentials such as committing access keys, passwords, and secrets to source control repositories. A compromised user's endpoint devices such as notebook, desktop, and mobile device will also allow an attacker to find secretive credentials. MITRE has listed a number of credentials dumping methods that can be exploited by attackers \cite{mitre2021c}. 

\subsection{Implement Network Controls}

In order to prevent DNS spoofing attacks, it is a good practice not to use Docker's default bridge docker0 but to use Docker's user-defined network \cite{cis2021a}. The developer using the end point device should encrypt the network with a virtual private network (VPN) and to regularly flush the device's DNS cache \cite{kaspersky2021a}. The VPN is also important to secure the communication between the containers \cite{goethals2019a}. %However, VPN can increase network latency and introduces delays that are bad for ad-hoc transient container applications such as event-triggered serverless functions or Internet of Things (IoT) containers communicating many small packets rapidly. Therefore, additional research is needed in network protection for such use-case. 
To protect the network connectivity from DoS attack, it is a good practice to turn on the intrusion detection and prevention systems (IDS and IPS) to detect and prevent such attacks. Lastly, it is recommended not exposing the Docker daemon socket (the main entry point for Docker API) \cite{owasp2021a} and other unnecessary ports (e.g., SSH Port 22).

\noindent \textbf{\textit{Limitations:}}
The use of VPN can increase network latency and introduces delays that are bad for ad-hoc transient container applications such as event-triggered serverless functions or Internet of Things (IoT) containers communicating many small packets rapidly. Therefore, additional research is needed in network protection for such use-case. IDS and IPS use rule or signature-based packet evaluation and therefore not effective against unknown attacks or against an attacker that poses as admin to "legitimately" log into the system \cite{dwyer2018a}. IDS which yields many false alarms can lead to "alert fatigue" while IPS can consume much network bandwidths.

\subsection{Robust Log Monitoring}
The mitigation measures need to enable the logging system to be robust and immutable. One method is the use of message authentication codes (MACs) and digital signatures to produce the secure logs, and to apply Bitcoin blockchain technique to produce a distributed log immutabilization solution \cite{cucurull2016a}, thus ensuring the logs' authenticity and non-repudiation. %However, there is transaction fees (at 0.00016 BTC/KB or USD6.83/KB as of 25th Sep 2021 \footnote{https://bitinfocharts.com/comparison/bitcoin-transactionfees.html}) when running the blockchain operation and is not sustainable in the long run.
To resolve the log storage problem, one practice is to use logging drivers to read the data directly from the Docker container's stdout and stderr ouput and to forward the logs to host machine or other endpoints such as syslog, journald, gelf, and others \cite{solarwinds2021a}. %Some limitations include the capacity limit of the local storage will determine the size of the log file \cite{docker2021g}. If the logs are sent remotely, a network failure will cause the lost of the logs \cite{docker2021o}.

\noindent \textbf{\textit{Limitations:}}
When running the blockchain operation, there is transaction fees (at 0.00016 BTC/KB or USD6.83/KB as of 25th Sep 2021\footnote{https://bitinfocharts.com/comparison/bitcoin-transactionfees.html}) and is not sustainable in the long run. Other limitations when using the logging drivers are that the capacity limit of the local storage will determine the size of the log file \cite{docker2021g}. If the logs are sent remotely, a network failure will cause the lost of the logs \cite{docker2021o}.

%\section{Discussion and Summary of Results}\label{discussion}
%\section{Discussion, Open Issues and Future Research Directions}\label{discussion} 
\section{Summary of Results and Future Research Directions}\label{summary} 
%
%The summary of STRIDE analysis of the components is listed in Table \ref{tab1}. 
% }

\subsection{Summary of Results}

The overall containers security analysis we conducted using the STRIDE framework is summarized in Table \ref{tab1}. It is observed that each of the STRIDE threat occurs in several DFD elements and results in multiple consequences with the aim to deceive, disrupt, disclose information, or to usurp control of the system. Spoofing is about using a fake identity to gain access into the system. GitHub (DS-1), Docker Hub (DS-2) and the containers (P-4, P-5) are the obvious targets for attackers to exploit and to introduce malicious contents in order to deceive (TC-2), retrieve info (TC-1) and to control the systems (TC-4). The efficient and ease-of-use characteristics of the container systems turn out to be the vulnerabilities for the threat to be successful. The ease of access into the code repository and image registry, unrestricted push and pull of the images, and the efficient sharing of host resources by several co-locating containers become the vulnerabilities.\\
%The protection measures are to enforce strict access controls, strengthen the image identity, perform rigorous threats scan, and to reduce systems capabilities in the containers.\\
\indent Tampering aims to modify the system or data with the intention to deceive (TC-1) the victim, steals the info (TC-1), disrupts the service (TC-3), and to gain control of the system (TC-4) via the tainted images. This threat has the widest impact to the DFD elements including the data stores of DS-1 and DS-2, all the data flow (DF) links, and the process of image build (P-2). In addition to the vulnerabilities listed earlier, the lack of container image governance is another vulnerability. Docker Hub is an open registry which is accessible by a private (paid membership) or community user. The images are freely uploaded and stored with no patch management or threats scanning rigor. Its integration into the automated CI/CD pipeline process further increases the attack surface.\\
\indent Repudiation occurs when an attacker denies an action which he has performed. The logs of a container is not stored in itself as the container is stateless and therefore the kernel will store the logs in the host storage (vulnerability V9). Due to the shared resources characteristic of co-locating containers (V12), an attacker can use a compromised container to access the kernel (P-7) to disable, modify or overwrite logs at the host storage.\\
\indent Information disclosure causes information to be revealed to attackers. The attackers will attempt to gain access to data stores at GitHub (DS-1) and Docker Hub (DS-2) to steal information about accounts, source codes, sensitive data, configuration files, etc. A skilled attacker can exploit the sensitive parameters used during the container configuration (P-3) to gain access to files in the host. He can also use the common shared network at the kernel (P-7) to connect two co-locating containers and to exchange unauthorised information.\\
\indent Denial of service (DoS) makes the system inaccessible for use. DoS can occur when an attack happens at each of the connecting "pipe" (DF-1,2,3,4,5) that links the elements in the container DFD system. A breakage in a connection will result in a change or patch in the application code not being updated in the final image and not deployed or updated in the application container. Proven tactics targeted at the resource isolation measures in the kernel can cause the host resources (eg. CPU, storage) to be inaccessible.\\
\indent Elevation of privilege grants the attacker access and control of the system. This is a serious threat which allows the attacker to take control (TC-4) of the host and carry out further damages. The tight integration of the container with the Linux kernel is a critical vulnerability (V10, V11, V12). Therefore, an attacker with access to a compromised container can utilize the Docker daemon (P-6) via exposed network ports and privilege system calls to attack the kernel (P-7) to obtain root control of the host.\\
%====================================================================================================
%\section{Open Issues and Future Research Directions}\label{future_irections}
%
\subsection{Future Research Directions}

Based on the above analysis, there are some areas which are open for further research. In our STRIDE threat modeling exercise, we focus on the “supply chain” from the code repository (using GitHub), to the image registry (Docker Hub), and finally to the Docker host with emphasis on the six elements of STRIDE (Spoofing, Tampering, Repudiation, Information Disclosure, Denial of Service, and Elevation of Privilege). Below, we outline some future research directions in containers security. 
%However, the wider and more holistic container ecosystem is connected and overlapped with the cloud and IoT ecosystems. Containers are used to build the cloud and IoT systems, and at the same time the cloud and IoT are using containers to run applications \cite{syed2018a}.

%\indent {\bf Vulnerabilities in IoT containers} Gartner predicts that the number of IoT devices will double every five years and it will reach 15 billion IoT devices by 2029, and they pose security risks to the enterprise infrastructures \cite{costello2021a}.Therefore, this is an important field to study the expanded attack vectors presented by the relationships between the container, cloud and IoT systems.\\
\indent {\bf Enhancement of container engine security} In this paper, we use Docker as the representative container engine for security survey as it is the most popular and pervasively used by enterprises and businesses. However, a couple of reports state that an alternative container engine called Kata\footnote{https://katacontainers.io/} container which is developed by IBM and Hyper.sh  can offer better security isolation while maintaining efficiency and performance and it has a strong reference customer in the form of Baidu AI Cloud \cite{li2020a}, \cite{flauzac2020a}, \cite{kumar2020a}. Therefore, another direction of study is a comprehensive comparison of the security and performance between Kata container and Docker container and investigate the possibility of a Docker substitute or areas for Docker's security enhancements.\\
\indent {\bf Security of alternative container technology} In recent years, there have been studies on Unikernel and its advantages of small footprint, speed and a reduced attack surface \cite{chen2021a},\cite{kuo2020a},\cite{olivier2019a},\cite{bratterud2017a},\cite{bratterud2015a}. This technology presents a useful area of study to determine the feasibility of replacing the container technology in order to reduce the vulnerabilities faced by the current container technology.\\
\indent {\bf Vulnerabilities of containers using different kernels} There are no comparison studies of container security between one which is based on Linux vs one based on Windows. Both the Linux and Windows kernels are designed differently and there is a large Windows application installed base and therefore it is of interest to know the comparative security strengths and weaknesses between the two. So far, most of the security analysis of the Windows and Linux operating systems were carried out several years ago and were considered out-dated \cite{zeng2016a},\cite{thomeczek2015a},\cite{salah2013a},\cite{bassil2012a},\cite{zhang2005a}.\\
\indent {\bf Evaluation of container scanning tools} There is little study about container vulnerability and threats detection tools and the evaluations of their performances. To date, there are many container image scanning tools such as Clair, Anchore, Trivy, etc. \cite{doerrfeld2021a} but few research into their effectiveness, their gaps and their impacts to the container's security. Javed and Toor of \cite{javed2021a} evaluated three scanners of Clair, Anchore and Microscanner in terms of the detection coverage and detection hit ratio for only 59 Docker Java-based images. Tunde-Onadele et al. \cite{onadele2019a} compared the detection accuracy of a static scanner (Clair) and a dynamic runtime detection scheme which analyzed the system call features using machine learning methods, like K-means, Self-Organizing Map and others to detect anomaly. Therefore, there is a need to study the available vulnerability detection methods and tools and to carry out a comprehensive evaluation of them.\\
%\indent {\bf Vulnerabilities relevant to language-based application containers} The top three programming languages used in Docker container application images in 2020 are Python, Go and Javascript and they have been growing steadily since 2015 \cite{lin2020b}. There is no known study of the types of threats and vulnerabilities that occur in the container and their attribution to the programming languages. Therefore, it is useful to the industry for researches to be carried out on the vulnerabilities of the different language-based applications when deployed in a container architecture.\\
\indent {\bf An end-to-end practical guide to securing containers} There is no one structured and integrated approach for container security. Today, each security tool or process only targets a specific area and to address it independently. For example, in the code build phase a developer will need to remember to scan the image, keep credentials in the secret vaults, verify the image signature and to sign it when pushing it to the registry, and all of these steps require different tools and processes. During the pull and deployment phase, a developer will need to scan the image for new vulnerabilities and to configure least privileges, network segmentation and least kernel interaction (e.g., minimal system calls) in runtime. The developer will then need to ensure the integrity of the images (e.g., patch and re-image) throughout the lifecycle of the container and to run monitoring and logging mechanisms to keep the container and its users safe. The National Institute of Standards and Technology (NIST) published a comprehensive container security guide in 2017 \cite{souppaya2017a} and it contained recommendations of best practices for specific components in a container architecture but did not provide working level details and its application in practical use-cases (e.g., via code repo, image registry, deployment, etc). Therefore, there is a need for the research community to produce industry relevant and practical guides for container security.

%===========================================================================================================================
%
%\begingroup
%\onecolumn
%\begin{multicols*}{1}
%\pagebreak
\begin{center}
{
\small
\renewcommand{\arraystretch}{2}%
%\begin{longtable}{|p{1.5cm}|p{1.5cm}|p{1.5cm}|p{3.5cm}|p{1.5cm}|p{3.5cm}|p{2.5cm}|}
\setlength\tabcolsep{5pt}
%\extracolsep{\fill}
\setlength\LTleft{0pt}
\setlength\LTright{0pt}
\begin{longtable}{|p{1.7cm}|p{1.5cm}|p{1.2cm}|p{3cm}|p{1.3cm}|p{3cm}|p{2.5cm}|}
\caption{Summary of our STRIDE Analysis} \label{tab1} \\
\hline
\textbf{STRIDE} & \textbf{Affected DFD Com\-p\-o\-n\-e\-n\-t\-s} & \textbf{Vulner\-a\-b\-i\-l\-i\-t\-i\-e\-s} & \textbf{Threat Actions} & \textbf{Threat Consequences} & \textbf{The Existing Mitigation Strategies} & \textbf{Limitations of the Mitigation Strategies} \\ \hline
\hline
%---------------------------------------------------------------------------
\multirow{4}{*}{\parbox{1.3cm}{Spoofing}}   & DS-1      & V1     & Spoof Github account by stealing credentials to gain access to GitHub account and to upload malicious codes.  & TC-1, TC-2  & MFA to protect account, scan image for vulnerabilities. & 2FA increases time and cost, SMS verification code can be attacked, and one-time passcodes generated by phone authenticator app can be stolen.   \\ \cline{2-7} 
& DS-1, DS-2        & V2    & Spoof GitHub or Docker Hub by using DNS hijack \& others.    & TC-2   & Protect network, use VPN, sign image, and scan image. & VPN introduces delay, and private keys for signature can be stolen.   \\ \cline{2-7} 
& DS-2  & V2  & Spoofing of Docker Hub account and image by exploiting typo squatting and "almost-similar name" image.     & TC-2, TC-4   & Use official or verified image, and scan image before upload to registry or download for deployment. & Container scanner is not foolproof and 34\% of vulnerabilities are undetected.  \\ \cline{2-7} 
& P-4, P-5      & V12   & Spoofing of DNS responses to all the container applications running on the Kubernetes cluster and to execute MITM attack on the network traffic between the containers. & TC-2            & Protect network, use VPN, and limit capabilities permissions in container (e.g. NET\_RAW). &  Same as above. \\ \hline
%-------------------------------------------------------------------------
\multirow{4}{*}{\parbox{1.3cm}{Tampering}}              & DF-4, DF-5   & V5    & MITM by attacker to insert malicious image in the connection between Docker Hub and the Docker host.   & TC-1, TC-2, TC-3          & Encrypt network, check for signature in image, verify hash, and scan image. &  Private keys for signature can be stolen, and container scanners are not highly accurate.   \\ \cline{2-7} 
& DF-1, DF-2, DF-3, DF-4, DF-5 & V7  & During the auto CI/CD pipeline, attacker can insert tampered and malicious images into any stage of the pipeline.    & TC-1, TC-3          & Protect network pipeline, scan code/image at each stage, sign image and verify it during deployment. & Container scanners are not highly accurate.  \\ \cline{2-7} 
& DS-2      & V2, V4  & Images in Docker Hub being tampered after attackers hacked into accounts. A vulnerability in an image takes an average of 181 days for it to be fixed and an extra 422 days to be updated.                  & TC-1, TC-2          & Scan image, sign image, and verify it during deployment, verify hash, and regular patching of image.& Same as above. Patching is manual, no testing for app compatibility before patch.\\ \cline{2-7} 
& P-2       & V2, V3    & When the image is built, malicious commands are injected into the image or tampered libraries are used in the application.                                              & TC-1, TC-2, TC-4    & Same as above. Keep image minimal, use multi-stage build, and use distroless images. & Same as above.   \\ \hline
%--------------------------------------------------------------------------------
\multirow{2}{*}{\parbox{1.3cm}{Repudiation}}            & P-7                           & V9, V10         & Disable logging, and modify logs.  & TC-2, TC-3          & Use message authentication code (MAC) and signature. Apply blockchain to distribute and immutabilize logs. & Transaction fees in blockchain is costly in the long term.   \\ \cline{2-7} 
& P-7                           & V9, V10         & Overwrite log disk space with junk.  & TC-2                & Use log drivers to store logs locally or to remote endpoints. & Local storage limit log size, and network failure causes logs to be lost.   \\ \hline
%--------------------------------------------------------------------------------
\multirow{3}{*}{\parbox{1.3cm}{Information Disclosure}} & DS-1, DS-2  & V1      & API and identity keys are exposed for attacker to take control of accounts in Github and Docker Hub.     & TC-1                & Do not store credentials and secrets in clear, keep them in "vaults". Use .gitignore to avoid uploading sensitive info during commit. & Credentials in "vaults" and compromised endpoints can be stolen, and user's negligence in sharing credentials   \\ \cline{2-7}  
& P-3                          & V10, V12        & Include sensitive parameters in the run command when deploying container.  & TC-1, TC-4          & Exercise diligence in not exposing sensitive parameters, scan for sensitive parameters and to raise alerts. & No significant limitation is reported or observed.  \\ \cline{2-7} 
& P-7                          & V10, V12        & Leakage of information between containers on the same host.  & TC-1                & Same as above. Harden host configuration, limit file permissions, and configure TLS for connections. & No significant limitation is reported or observed.  \\ \hline
%--------------------------------------------------------------------------------
\multirow{2}{*}{\parbox{1.3cm}{Denial of Service}}      & P-6, P-7           & V10             & Service disruption at the Host via kernel due to exception handling, disk write-back, logging, and others.                                                                   & TC-3     & Hardened configuration of cgroups to limit host resources usages eg. read-only filesystems, limit kernel calls, limit network communications, and use memory management module like MemGuard.   &  Cgroups and namespaces isolation are subjected to container escape and network attacks.                                                                         \\ \cline{2-7} 
& DF-1, DF-2, DF-3, DF-4, DF-5 & V7              & Any of the data flow pipes become disrupted to perform transmission of codes/images.  & TC-3                &  Install intrusion detection (IDS) and prevention systems (IPS) to protect the network connectivity.  & IDS can result in "alert fatigue" and IPS takes up network bandwidth.\\ \hline
%-------------------------------------------------------------------------------
\multirow{4}{*}{\parbox{1.3cm}{Elevation of Privilege}} & P-2       & V10, V11, V12   & Run container as root when it is not necessary. & TC-4                & Harden container configuration to just-enough privileges or run as "non-root mode". & No significant limitation is reported or observed.    \\ \cline{2-7} 
& P-6                          & V10, V11, V12   & Misconfiguration with network ports open.  & TC-4                & Scan network ports, do not expose Docker daemon socket and other unnecessary ports.  & No significant limitation is reported or observed. \\ \cline{2-7} 
& P-6                          & V10, V11, V12   & Enabling excessive  systems calls & TC-4                & Trace system calls and reduce unnecessary ones, analyse systems calls traffic and use Machine Learning techniques to detect anomaly, apply principle of privilege separation and partition container to smaller isolating containers. & Current ML techniques are not highly accurate and not tested for most use-cases. \\ \cline{2-7} 
& P-7                          & V10, V11, V12   & Memory attack by overcoming security of Linux and using TOCTOU techniques. & TC-4   & Same as above.    &    Same as above     \\ \hline
\end{longtable}
}
\end{center}
%\end{multicols*}
%\endgroup
%\twocolumn
% =======================================================================================================
\section{Conclusion}\label{conclusion}

The advancement of containers has helped enterprises and organizations to improve their processes and enable new business models. However, its full utilization has been daunted by the various security risks posed in the containers ecosystem. In this paper, we first assessed the security landscape in containers. In particular, we used the STRIDE framework to %\revised{survey the existing works in the research and practising communities in order to} 
identify vulnerabilities, threats and threat consequences on the entire container ecosystem. %, comprising the code commit, code repository GitHub, image build, image registry Docker Hub, image pull, and the containers. 
From our study, we found that many of the vulnerabilities are due to the containers' shared access to the host operating system's kernel. While there were isolation measures (e.g., namespaces) and resource control mechanisms (e.g., cgroups) in place, these could be breached when misconfigurations and liberal use of system calls and capabilities happened. From the ecosystem perspective, the numerous external entities who involved in writing the code, building the image, configuring the installation, setting up the network connectivities, and eventually deploying the application in production containers greatly increased the attack surfaces.

Then, we conducted a systematic survey on the existing works on containers security. In particular, we assessed the strengths and weaknesses of existing mitigation strategies against the identified security threats in containers. Based on our assessment, most of the existing mitigation strategies have certain limitations and not sufficient to address the security risks posed to the container systems. Therefore, we have also outlined several areas of future research directions % which even include alternative container technologies, such as Kata and Unikernel, 
to enhance the security of containers. We hope this paper will help practitioners and researchers to be aware of the current threat landscape and security gaps in containers, and open up areas for further explorations and studies. 

%Finally, we proposed some mitigation actions and we found that these were not exhaustive and they opened up areas for further research and explorations. 

%However, the benefits of using containers outweigh the risks of attacks and this would spur the continual usage of container for rapid and agile application development and deployment. 
%Therefore, we have outlined several areas of future research directions which even include alternative container technologies, such as Kata and Unikernel. We hope this paper will help practitioners and researchers to be aware of the security gaps existed in container systems, and will open up areas for further explorations and studies. 

% ======================================================================================================

%%
%% The acknowledgments section .
\section*{Acknowledgment}
We would like to thank our peers and colleagues for their valuable feedback. Any opinions, conclusions or recommendations expressed in this paper are those of the authors and do not necessarily reflect the views of the universities.

%\bibliographystyle{ACM-Reference-Format}
\bibliographystyle{splncs04}
\bibliography{container}

\end{document}